\documentclass[letterpaper,11pt]{article}
\usepackage{jheppub}

\usepackage{multirow}

\usepackage{subfig}
\usepackage{xspace}
\usepackage{xcolor}
\usepackage[countmax]{subfloat}
\usepackage{amsthm}

\usepackage{amsmath}

\DeclareMathOperator*{\argmin}{arg\,min}

\usepackage{color}
\DeclareRobustCommand{\Sec}[1]{Sec.~\ref{#1}}
\DeclareRobustCommand{\Secs}[2]{Secs.~\ref{#1} and \ref{#2}}
\DeclareRobustCommand{\App}[1]{App.~\ref{#1}}

\DeclareRobustCommand{\Fig}[1]{Fig.~\ref{#1}}

\DeclareRobustCommand{\Eq}[1]{Eq.~(\ref{#1})}
\DeclareRobustCommand{\Eqs}[2]{Eqs.~(\ref{#1}) and (\ref{#2})}
\DeclareRobustCommand{\InRef}[1]{Ref.~\cite{#1}}
\DeclareRobustCommand{\Refs}[1]{Refs.~\cite{#1}}

\preprint{MIT-CTP 5556}

\title{
A Spectral Metric for Collider Geometry
}

\author[a]{Andrew J. Larkoski}
\author[b,c]{and Jesse Thaler}

\affiliation[a]{Department of Physics and Astronomy, University of California, Los Angeles, CA 90095, USA}
\affiliation[b]{Center for Theoretical Physics, Massachusetts Institute of Technology, Cambridge, MA 02139, USA}
\affiliation[c]{The NSF AI Institute for Artificial Intelligence and Fundamental Interactions}
\emailAdd{larkoski@ucla.edu}
\emailAdd{jthaler@mit.edu}

\abstract{
By quantifying the distance between two collider events, one can triangulate a metric space and reframe collider data analysis as computational geometry.
One popular geometric approach is to first represent events as an energy flow on an idealized celestial sphere and then define the metric in terms of optimal transport in two dimensions.
In this paper, we advocate for representing events in terms of a spectral function that encodes pairwise particle angles and products of particle energies, which enables a metric distance defined in terms of one-dimensional optimal transport.
This approach has the advantage of automatically incorporating obvious isometries of the data, like rotations about the colliding beam axis.
It also facilitates first-principles calculations, since there are simple closed-form expressions for optimal transport in one dimension.
Up to isometries and event sets of measure zero, the spectral representation is unique, so the metric on the space of spectral functions is a metric on the space of events.
At lowest order in perturbation theory in electron-positron collisions, our metric is simply the summed squared invariant masses of the two event hemispheres.
Going to higher orders, we present predictions for the distribution of metric distances between jets in fixed-order and resummed perturbation theory as well as in parton-shower generators.
Finally, we speculate on whether the spectral approach could furnish a useful metric on the space of quantum field theories.
}

\begin{document}
\maketitle

\section{Introduction}

Events recorded in high-energy collider experiments populate an intricate and fascinating data manifold.
Na\"ively, this manifold has dimension $3N-4$, corresponding to the dimensionality of relativistic phase space for $N$ observed on-shell final-state particles.
For proton-proton collisions at the Large Hadron Collider (LHC), phase space can occupy hundreds of dimensions; for collisions of heavy ions, the phase space dimensionality can easily be in the thousands or even tens of thousands.
Such a large space is much too large to be interpreted and visualized directly, so most collider physics analyses involve some kind of data reduction.
Machine learning techniques have become increasingly popular in this context, since they can efficiently identify event structures according to the particular problem one wishes to solve; see e.g.~\Refs{Larkoski:2017jix,Guest:2018yhq,Albertsson:2018maf,radovic2018machine,Carleo:2019ptp,Bourilkov:2019yoi,Schwartz:2021ftp,Karagiorgi:2021ngt,Plehn:2022ftl} for recent reviews.

At the same time, there has been a growing interest in studying the manifold of collider data as a geometric object in its own right.
Considering the space of collider events as an abstract manifold means that one can evaluate quantities that encode properties of that manifold, such as its topology or local geometry. 
One approach to rigorously define a metric on the space of collider events is based on the energy mover's distance (EMD) \cite{Komiske:2019fks}, which quantifies the cost of rearranging the set of particles in one event to match the set of particles in another.
Computing the EMD involves solving an optimal transport problem for moving energy, analogous to the historical earth mover's distance that quantifies the minimal cost for rearranging piles of dirt~\cite{monge1781memoire} through the Wasserstein metric~\cite{kantorovich1939mathematical,wasserstein1969markov,dobrushin1970prescribing}.
The EMD has been used to define new collider observables~\cite{Komiske:2020qhg,Cesarotti:2020hwb,Ba:2023hix} and study the dimensionality of jets in CMS Open Data~\cite{Komiske:2019jim,Komiske:2022vxg}.
Variations on the EMD have been proposed for collider physics that simplify computations~\cite{Cai:2020vzx,Cai:2021hnn}, enable new physics searches~\cite{Mullin:2019mmh,CrispimRomao:2020ejk,Davis:2023lxq}, mitigate jet contamination~\cite{Alipour-fard:2023yjz}, improve computational efficiency~\cite{Tsan:2021brw,Kitouni:2022qyr}, and directly establish the Riemannian metric on phase space \cite{Larkoski:2020thc}.

In this paper, we propose an alternative metric for the space of collider events based on spectral functions.
With the original EMD, events are treated as distributions of energy on the celestial sphere where experimental calorimeters are located.
Because the celestial sphere is two dimensional, one has to solve a two-dimensional optimal transport problem, for which there are efficient algorithms~\cite{jacobi1890investigando,jacobi1890aequationum,kuhn1955hungarian} but no closed-form expressions.
Here, we introduce the \emph{spectral EMD}, where events are treated as one-dimensional spectral functions that encode the distribution of pairwise particle angles.
One-dimensional optimal transport has a simple expression in terms of quantile matching, and therefore our spectral EMD is more easily amenable to certain first-principles calculations.
Both the original EMD and the spectral EMD are infrared-and-collinear (IRC) safe, enabling studies of their properties in perturbative quantum chromodynamics (QCD).

As with many geometry problems, the choice of metric is a choice, which depends on the features one wants to expose about the data.
For example, the original EMD has the advantage of connecting to many well-known observables in collider physics, such as events shapes and jet substructure observables \cite{Komiske:2020qhg}.
It is, however, cumbersome to evaluate analytically in all but the most symmetric situations.
The spectral EMD has the advantage of automatically incorporating basic isometries of collider data, such as azimuthal rotations around the beam line.
Not every spectral function corresponds to physical collider event, though, so geodesics in spectral EMD space are more difficult to interpret.
We perform a few side-by-side comparisons of the original EMD and spectral EMD to emphasize these kinds of differences, using as our testbed the space of collimated sprays of particles called jets.

We emphasize that the spectral EMD is not just a one-dimensional projection of the original EMD.
Each one-dimensional projection of two-dimensional data involves some kind of information loss, whereas the spectral function preserves the complete event information, up to isometries and sets of measure zero.
By taking multiple one-dimensional projections, one can compute the sliced Wasserstein distance \cite{rabin2011wasserstein,bonneel2015sliced}, but this quantity behaves more like the original EMD than the spectral version.
In some ways, the spectral EMD behaves like the tangent earth mover's distance~\cite{10.1007/978-3-642-40020-9_43} in that both respect isometries.
Still, the original EMD and its variants have units of energy, whereas the spectral EMD has units of energy squared.

The rest of this paper is organized as follows.
In \Sec{sec:metric}, we review the spectral function and use it to define a metric distance between two events, up to isometries. 
In \Sec{sec:multcalc}, we evaluate the spectral metric in closed form between two jets with up to four particles in total, corresponding to a perturbative calculation through order $\alpha_s^2$.  
Using these results, we present the simplest, double-logarithmically accurate calculation of the distribution of metric distances in \Sec{sec:dlogsec}.
We find that the distance between quark/gluon jets is controlled by the sum of the corresponding quadratic color Casimirs, which was also observed in \InRef{Komiske:2022vxg} for the original EMD.
In \Sec{sec:fixedorder}, we numerically evaluate the distribution of the spectral EMD between jets at next-to-leading order in electron-positron collisions using the program EVENT2 \cite{Catani:1996vz} and note similarities of the results with non-global logarithms in the calculation of the light hemisphere mass \cite{Dasgupta:2001sh}.  
We generate quark and gluon jets using a parton shower in \Sec{sec:qgparton}, and find that the size of non-perturbative effects are smaller than what one might have expected from the analogy to angularities. 
In \Sec{sec:emdvsspec}, we present an analytical comparison between the spectral EMD and the original EMD for jets with low multiplicity, and present the first (to our knowledge) closed-form expressions for the EMD between two jets with up to two particles in them.
The spectral function philosophy enables the construction of metrics for more general spaces, and in \Sec{sec:thspacemet} we construct a metric on the space of quantum field theories that shares some features with the Zamolodchikov metric \cite{Zamolodchikov:1986gt,Kutasov:1988xb}.
We conclude in \Sec{sec:concs}, and look forward to numerous ways that further investigations into the spectral EMD could illuminate collider physics.

\section{The Spectral Metric}\label{sec:metric}

In this section, we introduce and define the spectral EMD on the space of jets.
This requires defining the spectral function, which has long been studied in the collider literature~\cite{Basham:1978bw,Tkachov:1995kk,Jankowiak:2011qa,Chen:2020vvp,Chakraborty:2019imr}.
While we focus our discussion on applications to jets and their substructure, these definitions easily extend to complete sets of particles in an event and to weighted point clouds more generally.

\subsection{Review of the Original EMD}
\label{sec:originalEMD}

To put the spectral EMD in context, it is useful to review the formulation of the original EMD~\cite{Komiske:2019fks}.
Consider a jet consisting of $N$ particles labeled by $i$, with energies $E_i$ and directions $\hat{n}_i$.
The energy flow of the jet is given by:
\begin{equation}
\label{eq:energyflow}
\mathcal{E}(\hat{n}) = \sum_{i=1}^N E_i \, \delta(\hat{n} - \hat{n}_i),
\end{equation}
which can be interpreted as the density of energy over an idealized detector at infinity with geometric coordinates $\hat{n}$.
Because of the inclusive sum over particles, the energy flow exhibits manifest invariance under the permutation group $S_N$.
The energy flow is normalized as:
\begin{equation}
\int d^2 \hat{n} \, \mathcal{E}(\hat{n}) = E_{\rm tot},
\end{equation}
where $E_\text{tot}$ is the total energy of the jet.
For hadron collider applications, one would typically replace energy with transverse momentum ($p_T$), but we stick with energy for our discussion for notational simplicity.

Given two energy flows ${\cal E}_A$ and ${\cal E}_B$, one can compute the optimal transportation cost between them~\cite{Komiske:2019fks}:
\begin{align}
\label{eq:original_EMD}
\text{EMD}_{\beta,R}({\cal E}_A,{\cal E}_B) = \min_{\{f_{ab}\}} \sum_{a \in J_A} \sum_{b \in J_B} f_{ab}\, \frac{\Omega(\hat{n}_a,\hat{n}_b)^\beta}{R^\beta} + \bigg| \sum_{a \in J_A} E_a - \sum_{b \in J_B} E_b  \bigg| .
\end{align}
The subscripts $a$ and $b$ denote particles in jets $A$ and $B$, respectively, $\Omega(\hat{n}_a,\hat{n}_b)$ is a pairwise angular distance between particles, $\beta \ge 1$ is an angular exponent, and $R$ is a fixed angular scale.
The energy transportation plan $f_{ab}$ satisfies the following inequalities:
\begin{align}
\label{eq:original_EMD_constraints}
f_{ab}\geq 0\,,\qquad \sum_{b \in J_B}f_{ab}\leq E_a\,,\qquad \sum_{a \in J_A}f_{ab}\leq E_b\,,\qquad \sum_{a \in J_A} \sum_{b \in J_B} f_{ab} = \min \Big(\sum_{a \in J_A} E_a, \sum_{b \in J_B} E_b \Big) \,.
\end{align}
As long as $R$ is larger than half the maximum distance between particles, this is a (modified) metric that satisfies the (modified) triangle inequality:
\begin{equation}
0 \le \text{EMD}_{\beta,R}({\cal E}_A,{\cal E}_B)^{1/\beta} \le \text{EMD}_{\beta,R}({\cal E}_A,{\cal E}_C)^{1/\beta} + \text{EMD}_{\beta,R}({\cal E}_B,{\cal E}_C)^{1/\beta}.
\end{equation}
If ${\cal E}_A$ and ${\cal E}_B$ have the same total energy, then $\text{EMD}_{\beta,R}({\cal E}_A,{\cal E}_B)^{1/\beta}$ is equivalent to the $p$-Wasserstein metric with $p = \beta$.
For $N$ particles per jet, a generic EMD solver requires $O(N^3 \log N)$ runtime.

The angular distance $\Omega(\hat{n}_a,\hat{n}_b)$ is also known as the ground metric, which has to satisfy its own triangle inequality:
\begin{equation}
0 \le \Omega(\hat{n}_a,\hat{n}_b) \le \Omega(\hat{n}_a,\hat{n}_c) + \Omega(\hat{n}_b,\hat{n}_c).
\end{equation}
For our calculations in $e^+ e^-$ collisions, we focus on $\beta = 2$, $R = 1$, and
\begin{equation}
\label{eq:capital_omega_def}
\Omega(\hat{n}_a,\hat{n}_b) \equiv \Omega_{ab} = 2 \sin \frac{\theta_{ab}}{2},
\end{equation}
where $\theta_{ab}$ is the opening angle between particles $a$ and $b$.
Note that with this normalization, $\Omega \in [0,2]$.
As discussed in \InRef{Ba:2023hix}, the EMD ``faithfully'' lifts the ground metric, meaning that if ${\cal E}_A$ and ${\cal E}_B$ are related by a translation of size $\Omega_0$, then the EMD is equal to $\Omega_0^\beta / R^\beta$.

\subsection{The Spectral Function and its Properties}
\label{sec:spectralfunctionproperties}

In this paper, we focus on an alternative way to represent a jet of $N$ particles via its spectral function:
\begin{align}
\label{eq:specfunc}
s(\omega) &= \sum_{i = 1}^N \sum_{j = 1}^N E_i\, E_j \,\delta \big(\omega-\omega(\hat{n}_i,\hat{n}_j) \big)\\
&= \sum_{i \in J} E_i^2 \, \delta(\omega) + 2 \sum_{i<j \in J} E_i \, E_j\,\delta \big(\omega-\omega(\hat{n}_i,\hat{n}_j) \big).\nonumber
\end{align}
Here, $\omega(\hat{n}_i,\hat{n}_j)$ is a pairwise angular distance between particles, which may or may not be related to $\Omega(\hat{n}_a,\hat{n}_b)$ above.
Because of the inclusive double sum over particles, the spectral function exhibits manifest invariance under the permutation group $S_N$.
Because the spectral function depends on pairwise distances, it is invariant to all isometries respected by $\omega$.
The spectral function is normalized as
\begin{align}
\int  d \omega \, s(\omega) = E^2_\text{tot}\,.
\end{align}
Like the energy flow in \Eq{eq:energyflow}, the spectral function in \Eq{eq:specfunc} is IRC safe since it exhibits (multi-)linear energy weighting, an inclusive sum over all particles, and only angular dependence inside the $\delta$-function.

The spectral function has a long history in collider and jet physics, starting with the energy-energy correlator \cite{Basham:1978bw}.
It has been used to represent the elements of a complete basis of IRC-safe observables \cite{Tkachov:1995kk}, to define observables for jet classification \cite{Jankowiak:2011qa}, to form the foundation of higher-point energy correlators \cite{Chen:2020vvp}, and to encode a jet's information for machine learning applications \cite{Chakraborty:2019imr}.
To the best of our knowledge, the spectral function has not yet been used to define a metric distance between two jets.

It is worth clarifying the difference between the angular distance $\omega$ in the spectral function and the angular distance $\Omega$ in the original EMD.
In both cases, we would like the pairwise distance to respect the isometries of the detector, which is $\text{O}(3)$ for the case of $e^+e^-$ colliders with spherical geometry.
The function $\omega$ in the spectral function is the pairwise distance between particles in the \emph{same} event.
With an appropriate choice of $\omega$, the spectral function of an individual jet is automatically invariant to isometries, as is any quantity defined in terms of spectral functions, including the spectral EMD defined below.
In particular, the spectral EMD between two jets that differ only by isometries is zero, which is often a desirable feature.
By contrast, the function $\Omega$ in the original EMD is the pairwise distance between particles in \emph{different} events.
The energy flow of an individual jet is not invariant to isometries, but with appropriate choice of $\Omega$, the original EMD between pairs of jets will respect isometries.
Crucially, the original EMD between two jets that differ only by isometries is \emph{not} zero.

For our studies, we use the angular measure:
\begin{equation}
\label{eq:lower_omega}
\omega(\hat{n}_i,\hat{n}_j) \equiv \omega_{ij} = \frac{\Omega(\hat{n}_i,\hat{n}_j)^\beta}{2}.
\end{equation}
This normalization has been chosen such that $\omega_{ij} = 1 - \cos \theta_{ij}$ for $\beta = 2$, which is commonly used in the spectral function literature.
Note, though, that the meaning of $\beta$ is different between the original EMD and the spectral function.
For the original EMD, $\beta$ changes the optimal transport problem between two jets.
For the spectral function, $\beta$ changes the representation of an individual jet, independent of optimal transport.
In general, $\omega$ need not be a ground metric satisfying a triangle inequality, though it typically will be, since isometries are defined as distance preserving maps between metric spaces.
Despite the potential for confusion, we will use the same symbol $\beta$ since this ensures that the original EMD and the spectral EMD will behave similarly in simple limits, as discussed further in \Sec{sec:emdvsspec}.

Because the spectral function is invariant to isometries, one cannot reconstruct an event uniquely from its spectral representation.
As shown in \App{app:uniqueness}, though, the spectral function does determine an event uniquely up to isometries and pathological cases of measure zero.
Theorem 2.6 of \InRef{BOUTIN2004709} proves that two point clouds in $\mathbb{R}^k$ are equal, up to the action of an arbitrary isometry group, if their distributions of pairwise distances are identical.
This proof can be lifted to weighted point clouds apart from special configurations with degenerate distances, which occupy a space of measure zero in phase space.
While the uniqueness of the spectral representation will not be needed for our analysis, we find it satisfying that the metric on the space of spectral functions furnishes a metric on the space of events, modulo isometries and measure zero regions of phase space.

\subsection{Introducing the Spectral EMD}

Following the same logic as in \Sec{sec:originalEMD}, we can define the distance between two spectral functions $s_A(\omega)$ and $s_B(\omega)$ as the minimal cost to rearrange the spectral functions to be identical.
Like with the original EMD, we have to choose a ground metric for $\omega$, which we always take to be the Euclidean distance $|\omega - \omega'|$.
Assuming a Euclidean ground metric, we can leverage the closed form expression for the 1-Wasserstein distance in one dimension.

The spectral EMD is defined as:
\begin{equation}
\label{eq:SEMD_def}
\text{SEMD}_{\beta}(s_A, s_B) \equiv \int_0^{\omega_{\rm max}} d \omega  \, \big| S_A(\omega) - S_B(\omega) \big| .
\end{equation}
The subscript $\beta$ refers to the angular measure in \Eq{eq:lower_omega}, which implies the maximum angular value $\omega_{\rm max} = 2^{\beta -1}$.
The spectral EMD depends on the cumulative spectral function $S(\omega_{\rm upper})$:
\begin{align}
S(\omega_{\rm upper}) \equiv \int_0^{\omega_{\rm upper}} d\omega \, s(\omega) = \sum_{i =  1}^N  \sum_{j=  1}^N E_i \, E_j\, \Theta\big(\omega_{\rm upper}-\omega(\hat{n}_i,\hat{n}_j) \big),
\end{align}
where $\Theta$ is the Heaviside function.
For $N$ particles per jet, computing the spectral EMD is dominated by computing the spectral function and sorting its entries to find the cumulative distribution, which takes $O(N^2 \log N)$ runtime (cf.~$O(N^3 \log N)$ for the original EMD).

Because $S_A(\omega_{\rm max}) = E^2_{A}$  and $S_B(\omega_{\rm max}) = E^2_{B}$ could be different, this is an example of an unbalanced transport problem.
It is straightforward to map this to a balanced transport problem.
Letting $E^2_{A} > E^2_{B}$ without loss of generality, we introduce a modified spectral function with a ``reservoir'' at $\omega_{\rm max}$:
\begin{equation}
\label{eq:balancing}
s_B^{\rm mod}(\omega) = s_B(\omega) + \big(E^2_{A} - E^2_{B} \big) \delta(\omega-\omega_{\rm max}),
\end{equation}
such that $s_A$ and $s_B^{\rm mod}$ have the same total weight.
This modification leaves \Eq{eq:SEMD_def} unchanged:
\begin{equation}
\text{SEMD}_{\beta}(s_A, s_B) = \text{SEMD}_{\beta}(s_A, s_B^{\rm mod}),
\end{equation}
which is identical to the 1-Wasserstein metric, up to an overall normalization factor of $E^2_{{\rm tot},A}$.

In the body of this paper, we restrict our attention to the 1-Wasserstein metric.
More generally, after using \Eq{eq:balancing} to make this a balanced transport problem with total weight $E^2_{\rm tot}$, one could consider the ($p$-th power of the) $p$-Wasserstein metric:
\begin{equation}
\label{eq:SEMD_p_def}
\text{SEMD}_{\beta,p}(s_A, s_B) \equiv \int_{0}^{E^2_{\rm tot}} dE^2\, \big| S^{-1}_A(E^2) - S^{-1}_B(E^2) \big|^p .
\end{equation}
This expression depends on the inverse of the cumulative spectral function $S^{-1}(E_{\rm upper}^2)$, which yields the value of $\omega_{\rm upper}$ that encloses $E_{\rm upper}^2$ of spectral weight.
For the special case of $p = 1$, \Eq{eq:SEMD_p_def} is equivalent to \Eq{eq:SEMD_def}, since both are expressions for the area between the two cumulative spectral functions.
In \App{sec:riemannspec}, we present some results for $p=2$.

On normalized probability distributions, the Wasserstein distance satisfies the properties of a metric: identity of indiscernibles, symmetry, and the triangle inequality.
Thus, after doing the weight balancing trick, our spectral EMD is indeed a metric on the space of spectral functions, $\{s(\omega)\}$.
As argued in \App{app:uniqueness}, the spectral function uniquely defines a jet up to isometries and sets of measure zero.
Therefore, the spectral EMD is also a metric distance on the space of jets, $\{J\}$.
That is, if the Wasserstein distance between two spectral functions $s_A(\omega)$ and $s_B(\omega)$ is 0, then the two jets $A$ and $B$ are identical, up to isometries and pathological configurations.

\subsection{Example Optimal Transport Plan}
\label{sec:example_ot}

To gain some intuition for the spectral EMD, it is useful to consider a low multiplicity example.
We will do a more systematic study in \Sec{sec:multcalc}, but here we identify some features of the optimal transport plan for spectral functions.

The optimal transport plan can be viewed as a ``geodesic'' between two spectral functions $s_A(\omega)$ and $s_B(\omega)$.
Introducing a ``time'' parameter $t\in[0,1]$, we can envision continuously transforming one spectral function into another with a minimal cost at each time step.
In general, this transportation plan is not unique, but there is a convenient constant speed geodesic for one-dimensional distributions:
\begin{align}
\label{eq:OTP}
S_\text{OTP}(\omega;t) = \left[ (1-t) \, S_A^{-1}+t \, S_B^{-1} \right]^{-1}(\omega)\,,
\end{align}
where $S^{-1}$ is the functional inverse of the cumulative spectral function.
By construction, $S_\text{OTP}(\omega;0) = S_A(\omega)$ and $S_\text{OTP}(\omega;1) = S_B(\omega)$.
To determine the optimal transport plan spectral function, we simply differentiate \Eq{eq:OTP} with respect to $\omega$:
\begin{equation}
\label{eq:OTP_deriv}
s_\text{OTP}(\omega;t) = \frac{d}{d\omega}S_\text{OTP}(\omega;t).
\end{equation}

\begin{figure}[t!]
\centering
\includegraphics[width=0.5\textwidth]{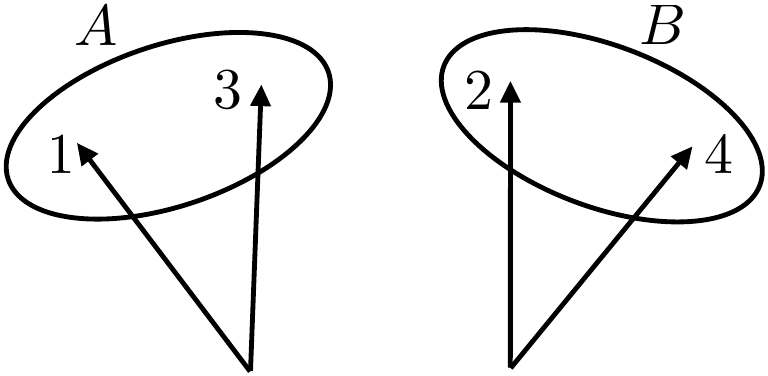}
\caption{
Labeling convention for particles in this paper.
Jet $A$ consists of odd-numbered particles and jet $B$ consists of even-numbered particles.
}
\label{fig:jetAB}
\end{figure}

As an example, consider the optimal transport plan between two jets that each contain two particles and have the same total energy $E$, as shown in \Fig{fig:jetAB}:
\begin{itemize}
\item Jet $A$ consists of odd-numbered particles $\{1,3\}$; and
\item Jet $B$ consists of even-numbered particles $\{2,4\}$.
\end{itemize}
Their cumulative spectral functions are:
\begin{align}
\label{eq:2particle_1}
S_A(\omega) &= \big(E_1^2+E_3^2 \big) \Theta(\omega) + 2E_1E_3\,\Theta \big(\omega-\omega_{13} \big),\\
\label{eq:2particle_2}
S_B(\omega) &= \big(E_2^2+E_4^2\big) \Theta(\omega) + 2E_2E_4\,\Theta\big(\omega-\omega_{24}\big).
\end{align}
The inverse cumulative spectral functions are then:
\begin{align}
\label{eq:inv_cum_A}
S_A^{-1}(x) &= \Theta\big(x-(E_1^2+E_3^2)\big)\,\omega_{13}+\Theta(x-E^2)\,(1-\omega_{13}),\\
\label{eq:inv_cum_B}
S_B^{-1}(x) &= \Theta\big(x-(E_2^2+E_4^2) \big)\, \omega_{24} + \Theta(x-E^2)\, (1-\omega_{24}).
\end{align}
The optimal transportation plan is simple enough that the inverses in \Eq{eq:OTP} can be taken analytically, as can the derivative in \Eq{eq:OTP_deriv}.
The final result for the optimal transport plan spectral function is:
\begin{align}
s_\text{OTP}(\omega;t)&= \Theta(E_2 E_4-E_1 E_3)\Big[
\delta(\omega)(
E_2^2+E_4^2
)+\delta(
\omega-t\,\omega_{24}
)(
2E_2E_4-2E_1E_3
)\\
&
\hspace{4cm}+\,\delta(
\omega-(1-t)\, \omega_{13} -t\,\omega_{24}
)\,2E_1 E_3 \Big]\nonumber\\
&\hspace{1cm}
+\Theta(E_1 E_3-E_2 E_4)\Big[
\delta(\omega)(
E_1^2+E_3^2
)+\delta(
\omega-(1-t)\, \omega_{13}
)(
2E_1E_3-2E_2E_4
) \nonumber\\
&
\hspace{4cm}+\,\delta(
\omega-(1-t)\,\omega_{13}-t\,\omega_{24}
)\,2E_2E_4\Big].\nonumber
\end{align}
As required, this spectral function reduces to $s_A(\omega)$ and $s_B(\omega)$ at $t=0$ and $t=1$, respectively.

\begin{figure}[t]
\begin{center}
\includegraphics[width=0.24\textwidth]{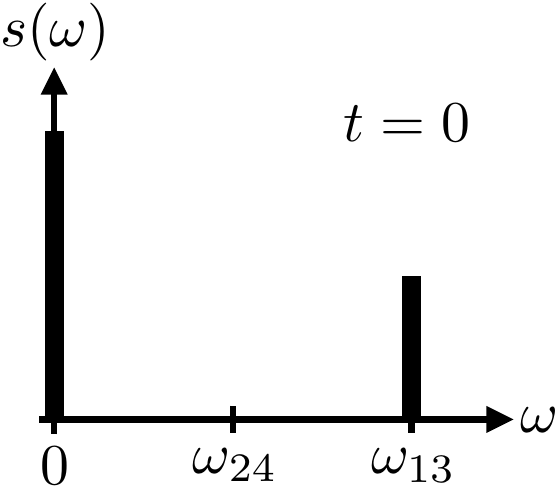} \includegraphics[width=0.24\textwidth]{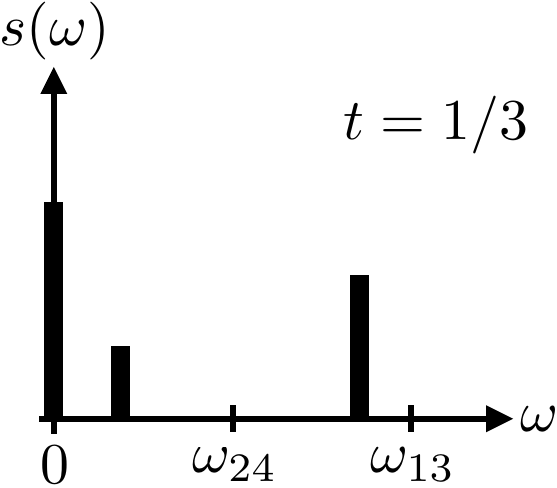} \includegraphics[width=0.24\textwidth]{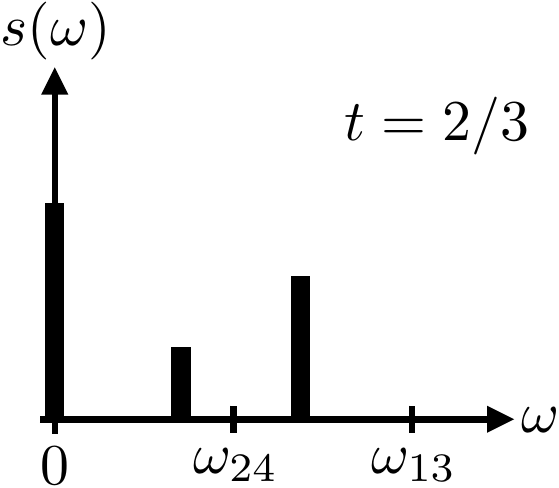} \includegraphics[width=0.24\textwidth]{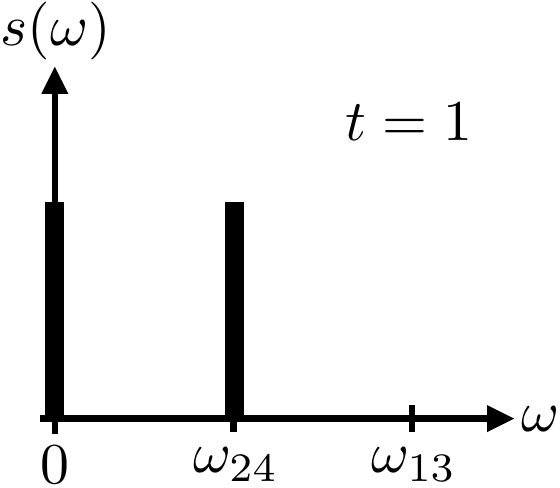}
\caption{\label{fig:otp_spec}
Snapshots of the optimal transport plan between two spectral functions.
}
\end{center}
\end{figure}

An example of this optimal transport plan is illustrated in \Fig{fig:otp_spec}, for the case of
\begin{equation}
E_1 E_3 = \frac{E^2}{6}, \quad E_2 E_4 = \frac{E^2}{4}, \quad \omega_{13} = \frac{2}{3}, \quad \omega_{24} = \frac{1}{3}.
\end{equation}
Note that the intermediate spectral functions at times $t\in(0,1)$ have two peaks away from $z=0$.
A collection of $n$ particles on the plane has ${n\choose 2}$ pairwise angles and of course there is no integer $n$ for which ${n\choose 2}=2$.
Thus, these intermediate spectral functions cannot be mapped to a generic collection of particles on the plane.
In this way, the optimal transport plan between spectral functions does not generically correspond to a rearrangement of the particles.

This behavior is distinct from original EMD behavior in \InRef{Komiske:2019fks}, where the optimal transportation plan directly measures the cost of rearranging particles on the plane.
Since the spectral function is blind to isometries, it is not surprising that the spectral optimal transportation plan does not have a particle interpretation.
In our view, this is not a problem but simply a choice.
What it does imply, though, is that the space of jets as defined by their spectral functions will have a different structure than the space of jets defined by particles.
We start to compare these two spaces on low-multiplicity jets in \Sec{sec:emdvsspec}, but leave a detailed study to future work.

\section{Spectral Metric Between Low-Multiplicity Jets}
\label{sec:multcalc}

In this section, we explicitly calculate the spectral metric between two jets with low constituent multiplicities.
This will concretely illustrate what information is encoded in the metric and explicitly demonstrate its IRC safety.
We arrange our analysis perturbatively in the strong coupling $\alpha_s$, and consider jets with up through three particles, corresponding to a relative ${\cal O}(\alpha_s^2)$ compared to leading order.
The following expressions will be used compute analytic distributions of the spectral metric in \Secs{sec:dlogsec}{sec:fixedorder}.

\subsection{${\cal O}(\alpha_s^0)$: Jets with One Particle}
\label{sec:one_particle}

At lowest order in an $\alpha_s$ expansion, the two jets being compared each consist of a single particle.
Hence, their spectral functions are:
\begin{align}
&s_A^{(0)}(\omega) = E_A^2\,\delta(\omega),&s^{(0)}_B(\omega) = E_B^2\,\delta(\omega),
\end{align}
where $E_A$ are $E_B$ are the respective energies of the two jets, and the superscript denotes the order in $\alpha_s$.
Using \Eq{eq:SEMD_def}, their spectral metric distance is 
\begin{align}
\label{eq:SEMD_00}
\text{SEMD}^{(0,0)}_{\beta}(s_A, s_B) = \int_0^{\omega_{\max}}  d\omega \, |E_A^2-E_B^2| = |E_A^2-E_B^2|\,\omega_{\max},
\end{align}
where the $(k,\ell)$ superscript means that we are working to order $\alpha_s^k$ for jet $A$ and order $\alpha_s^\ell$ for jet $B$. 
Framed as an optimal transport problem, this distance corresponds to the minimal energy-squared that must be eliminated to render the spectral functions identical.
Eliminating energy corresponds to transporting it ``out'' of the jet, from $\omega=0$ to immediately beyond $\omega=\omega_{\max}$, where the extra energy-squared can be dumped.
The distance that this extra energy must be carried is $\omega_{\max}$ in units of $\omega$ in this framework, and so the total cost of removing this energy is the distance carried times the amount of excess energy.

As long as $\omega_{\max}$ is larger than the maximum $\omega$ in either spectral functions, \Eq{eq:SEMD_def} defines a metric distance.
For two jets with radius $R$, it is natural, though not necessary, to set $\omega_{\max} \simeq R^\beta / 2$, such that the excess energy would be proportional to $|E_A^2-E_B^2|R^\beta$.
We prefer to leave $\omega_{\max}$ as a free parameter, since it enables a meaningful comparison of jets that differ both in total energy and in jet radius used to define them.

\subsection{${\cal O}(\alpha_s^1)$: Jets with Up to Two Particles}
\label{sec:two_particles}

The contribution to the spectral metric at ${\cal O}(\alpha_s^1)$ comes from two sources:
\begin{itemize}
\item Jet $A$ consists of two particles and jet $B$ consists of only one; or
\item Vice versa.
\end{itemize}
For simplicity, we assume that all final state particles are massless.
For jet $A$ consisting of two particles $\{1,3\}$, its spectral function is:
\begin{align}
\label{eq:specA_1}
s^{(1)}_A(\omega) = (E_1^2+E_3^2)\,\delta(\omega) + 2E_1E_3\,\delta\left(\omega-\omega_{13}\right),
\end{align}
where $E_1+E_3 = E_A$, the total energy of jet $A$.
This yields a contribution to the spectral metric between jets $A$ and $B$ of:
\begin{align}
\label{eq:metas}
\text{SEMD}^{(1,0)}_{\beta}(s_A, s_B) &= \int_0^{\omega_{\max}} d\omega\, \left|
E_1^2+E_3^2+2E_1E_3\,\Theta\left(\omega-\omega_{13}\right) - E_B^2
\right| \\
&=|E_1^2+E_3^2 - E_B^2|\, \omega_{13}+|E_A^2-E_B^2|\,(\omega_{\max}-\omega_{13}), \nonumber
\end{align}
where $\omega_{13}$ is the angular distance between particles $1$ and $3$.
The complete metric distance at this order in perturbation theory is the sum of this result with the corresponding configuration when jet $B$ consists of two particles and jet $A$ only has a single particle, i.e. $\text{SEMD}^{(0,1)}_{\beta}$.

To interpret this result a bit more clearly, let us assume that the jet energies are identical, $E_A=E_B\equiv E$, and use the canonical choice of $\beta = 2$.
Then, \Eq{eq:metas} simplifies to
\begin{align}
\label{eq:SEMD_mass}
\text{SEMD}^{(1,0)}_{\beta=2}(s_A, s_B) &= |E_1^2+E_3^2 - E^2|\, \omega_{13}=2E_1E_3\left( 1-\cos\theta_{13}\right)\\
&=m_A^2\,,\nonumber
\end{align}
which is just the squared mass of jet $A$.
The metric distance through this order is therefore:%
\footnote{Strictly speaking, one cannot simply add the contributions, since different orders appear in different parts of the calculation.  In this case, though, the mass is zero at lowest order, we can use the same observable for $\text{SEMD}^{(1,0)}$ and $\text{SEMD}^{(0,1)}$.}
\begin{align}
\text{SEMD}^{(1,0)}_{\beta=2} + \text{SEMD}^{(0,1)}_{\beta=2}  = m_A^2+m_B^2\,.
\end{align}
Because the jet mass is itself an IRC-safe observable,  the metric distance is too, at least through ${\cal O}(\alpha_s^1)$.

\subsection{${\cal O}(\alpha_s^2)$: Jets with Up to Three Particles}
\label{sec:uptothree}

To simplify the analysis and illustrate the unique features of the spectral metric at ${\cal O}(\alpha_s^2)$, we assume that jets $A$ and $B$ have the same energy $E_A=E_B\equiv E$.
Now, there are three possible configurations that must be considered:
\begin{itemize}
\item Jet $A$ consists of three particles (${\cal O}(\alpha_s^2)$) and jet $B$ has a single particle (${\cal O}(\alpha_s^0)$);
\item Vice versa; or 
\item Jets $A$ and $B$ both consist of two particles (two factors of ${\cal O}(\alpha_s^1)$).
\end{itemize}
We continue to use the convention that jet $A$ ($B$) consists of odd-numbered (even-numbered) particles.

\begin{figure}[t!]
\begin{center}
    \subfloat[]{
\includegraphics[width=0.38\textwidth]{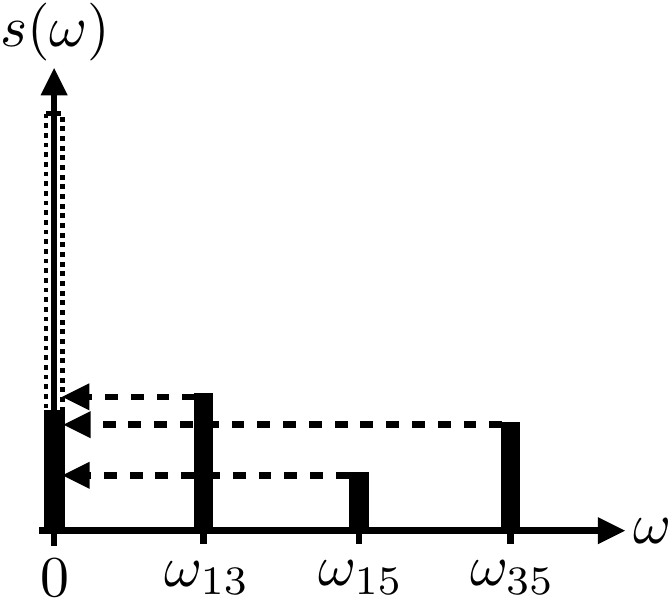}
\label{fig:3moves0_a}
}\ \ \ \ \ \ 
    \subfloat[]{
\includegraphics[width=0.38\textwidth]{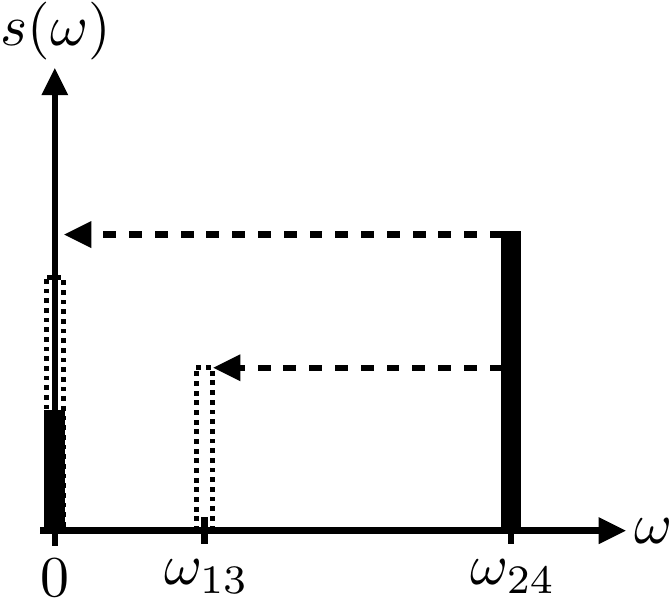}
\label{fig:3moves0_b}
}
\caption{\label{fig:3moves0}
Illustration of the optimal transportation plan between two jets at ${\cal O}(\alpha_s^2)$.
The heights of the peaks in the spectral functions are representative and sum to the same total value, and the locations of the peaks satisfy the Pythagorean theorem.
(a) Transport of three $\delta$-functions of the jet $A$ spectral function (solid) to render it identical to the jet $B$ spectral function (dashed), isolated to the origin.
(b) Transport between the spectral functions of two jets (solid and dashed), each with two constituent particles.
Part of the rightmost peak is transported to the origin and part is transported to the location of the peak in the other spectral function.
}
\end{center}
\end{figure}

We start with the first two configurations.
Considering jet $A$ with particles $\{1,3,5\}$, its spectral function is:
\begin{align}\label{eq:spec3mass}
s_A^{(2)}(\omega) = (E_1^2+E_3^2+E_5^2)\delta(\omega) &+ 2E_1E_3\,\delta(\omega-\omega_{13})\\
&+ 2E_1E_5\,\delta(\omega-\omega_{15})+ 2E_3E_5\,\delta(\omega-\omega_{35}),\nonumber
\end{align}
with $E_1+E_3+E_5 = E$.
At this order, the spectral function of jet $B$ is
\begin{align}
s^{(0)}_B(\omega) = E^2\,\delta(\omega)\,.
\end{align}
While it is straightforward to calculate the spectral distance from its integral representation, it is convenient to think in terms of the optimal transportation plan shown in \Fig{fig:3moves0_a}.
To make $s_A^{(2)}$ identical to $s_B^{(0)}$, we must transport each of the $\delta$-functions at $\omega > 0$ to $\omega=0$.
The cost of making these moves is the distance times the squared energy weight for each $\delta$-function, yielding a distance between spectral functions of:
\begin{align}
\label{eq:spectral20}
\text{SEMD}^{(2,0)}_\beta(s_A, s_B)&= 2E_1 E_3 \, \omega_{13}+2E_1E_5 \, \omega_{15}+2E_3E_5 \, \omega_{35}.
\end{align}
For the special case of $\beta = 2$, this is just the total squared mass of jet $A$.
Symmetrizing over the two jets, the $\beta = 2$ spectral distance to this order at least consists of
\begin{align}\label{eq:massmet3}
\text{SEMD}^{(2,0)}_{\beta=2}(s_A, s_B) + \text{SEMD}^{(0,2)}_{\beta=2}(s_A, s_B)&= m_A^2+m_B^2.
\end{align}

Next, consider the configuration for which both jets have two constituent particles.
This is the same configuration as in \Sec{sec:example_ot}, where jet $A$ has particles $\{1,3\}$ and jet $B$ has particles $\{2,4\}$.
The cumulative spectral functions were given already in \Eqs{eq:2particle_1}{eq:2particle_2}.
Using the integral representation in \Eq{eq:SEMD_def}, the spectral metric contains:
\begin{align}
\label{eq:angordmet3}
\text{SEMD}^{(1,1)}_{\beta} &= \int_0^{\omega_{\max}} d \omega \, \left|
E_1^2+E_3^2+2E_1E_3\,\Theta\left(
\omega-\omega_{13}
\right)-(E_2^2+E_4^2)-2E_2E_4\,\Theta\left(
\omega-\omega_{24}
\right)
\right|\nonumber\\
&=|2E_1E_3-2E_2E_4|\min\left[
\omega_{13},\omega_{24}
\right]+\Theta\left(
\omega_{13}-\omega_{24}
\right)\,2E_1E_3\left(
\omega_{13}-\omega_{24}
\right)\nonumber\\
&
\hspace{2cm}
+\Theta\left(
\omega_{24}-\omega_{13}
\right)\,2E_2 E_4\left(
\omega_{24}-\omega_{13}
\right)
.
\end{align}
Even with $\beta = 2$, this expression cannot be reduced to the jet mass.
Instead, it contains a detailed comparison between the angular separation and the product of energies of the particles in the jets, in a way that cannot be captured by mass alone.

An alternative way to derive \Eq{eq:angordmet3} is using the optimal transportation plan in \Fig{fig:3moves0_b}.
The less energetic $\delta$-function is moved to the location of the more energetic $\delta$-function, and then the extra energy of the more energetic $\delta$-function is moved to the origin.
The cost of these moves is:
\begin{align}\label{eq:enordmet3}
\text{SEMD}^{(1,1)}_{\beta}(s_A, s_B) &= \min[2E_1E_3,2E_2E_4]\left|\omega_{13}-\omega_{24}\right|\\
&
\qquad+\Theta(E_1E_3-E_2E_4)\,(2E_1E_3-2E_2E_4)\,\omega_{13}\nonumber\\
&
\qquad+\Theta(E_2E_4-E_1E_3)\,(2E_2E_4-2E_1E_3)\,\omega_{24}\,.\nonumber
\end{align}
By enumerating all cases, it is straightforward to verify that \Eqs{eq:angordmet3}{eq:enordmet3} are indeed identical.
These two methods of performing calculation illustrate distinct ways of thinking about the problem:  either ordering in the product of energies or in the pairwise angles.%
\footnote{This equivalence only holds for $p = 1$.  The method of constructing the optimal transport from energy ordering is more general, yielding the $p$-Wasserstein metric from \Eq{eq:SEMD_p_def}.
Part of the reason why we prefer $p =1$ is that energy and angular ordering are related by inverting the cumulative spectral function.
}

Combining the results of \Eqs{eq:massmet3}{eq:angordmet3}, the spectral metric between two jets at ${\cal O}(\alpha_s^2)$ takes the compact expression:
\begin{align}\label{eq:genmetas2}
\text{SEMD}_\beta^{(2)}(s_A, s_B) &= \sum_{a,a'\in J_A }E_aE_{a'}\omega_{aa'}+\sum_{b,b'\in J_B}E_bE_{b'}\omega_{bb'}\\
&
\qquad\qquad-\sum_{\substack{a,a'\in J_A \\b,b'\in J_B}}\min[E_aE_{a'},E_bE_{b'}]\min[\omega_{aa'},\omega_{bb'}]\,,\nonumber
\end{align}
where the sums run over all possible pairs of particles $a,a'$ in jet $A$ and $b,b'$ in jet $B$.
If any of the energies vanish or the particles in a jet become collinear, the distance reduces to an energy-energy correlation function of the more massive jet, making IRC safety manifest.
These calculations can be continued to higher orders, and one can observe how IRC safety manifests itself at every perturbative order, as it must, due to the IRC safety of the spectral functions themselves.

\section{Double-Logarithmic Distance Distributions Between Jets}
\label{sec:dlogsec}

We now move to calculating the distribution of distances between jets initiated by different partons.
The distribution of distances depends on the squared-amplitudes for two processes:
\begin{align}
\label{eq:varrho_def}
\frac{d\varrho}{d \ell_\beta}\equiv \frac{1}{Z} \int d\vec x_A\, d\vec x_B\, |{\cal M}_A|^2\,|{\cal M}_B|^2\,\delta\big(
\ell_\beta - \text{SEMD}_\beta(s_A, s_B)
\big).
\end{align}
Here, $\vec x_A$ is the vector of phase space coordinates for jet $A$ and ${\cal M}_B$ is the corresponding matrix element, and similarly for jet $B$.  
This quantity is \emph{not} a cross section, since the integration is over two phase spaces for distinct jets.
With the normalization factor $Z$, \Eq{eq:varrho_def} is nevertheless a probability density for $\ell_\beta$, which justifies the notation $\varrho$.

For the analyses in this paper, we focus on jets initiated by quarks or gluons, so the matrix elements can be calculated in perturbative QCD.  
Then, we can expand \Eq{eq:varrho_def} in powers of $\alpha_s$:
\begin{align}
\frac{d\varrho}{d\ell_\beta} = \frac{d\varrho^{(0)}}{d\ell_\beta}+\frac{\alpha_s}{2\pi}\,\frac{d\varrho^{(1)}}{d\ell_\beta}+\left(\frac{\alpha_s}{2\pi}\right)^2\frac{d\varrho^{(2)}}{d\ell_\beta}+\cdots\,.
\end{align}
We established in \Sec{sec:multcalc} that it is sufficient at ${\cal O}(\alpha_s^0)$ and ${\cal O}(\alpha_s^1)$ to let at least one of the jets be massless, such that the corresponding spectral function is simply a delta function at the origin.
Therefore, through ${\cal O}(\alpha_s^1)$, this distribution is equivalent to the distribution of an energy correlation function-like observable \cite{Banfi:2004yd,Larkoski:2013eya} measured on a single jet.  
Starting at ${\cal O}(\alpha_s^2)$, though, the distance $\ell_\beta$ describes honest correlations between jets that both have non-zero mass, as discussed further in \Sec{sec:fixedorder}.

It is straightforward to compute the distribution of distances at double-logarithmic accuracy, where jet emissions are strongly-ordered in both energy and angle.
At this accuracy, we can immediately write down resummed results for the distribution of distances between two jets, since the spectral EMD is dominated by a single emission.
Using \Eq{eq:metas} and taking the two jets to have equal energy $E$, the dimensionless distance at double-logarithmic accuracy is
\begin{align}
\label{eq:tilde_s}
\tilde \ell_\beta \equiv \frac{\ell_\beta}{E^2} \simeq z_A (1-z_A) \Omega_{A}^\beta + z_B (1-z_B) \Omega_{B}^\beta.
\end{align}
Here, the energy fractions and angles are:
\begin{align}
z_A = \frac{\min[E_1,E_3]}{E}, \quad \Omega_A = 2 \sin \frac{\theta_{13}}{2}, \quad z_B= \frac{\min[E_2,E_4]}{E}, \quad \Omega_B = 2 \sin \frac{\theta_{24}}{2}.
\end{align}
For strongly-ordered emissions, only one term in \Eq{eq:tilde_s} will dominate for a given phase space configuration.
Thus, to double-logarithmic accuracy, $\tilde \ell_\beta$ is simply the sum of two dimensionless energy-energy correlation functions with angular exponent $\beta$, and the distribution of this observable can therefore be copied from standard results.

The resummed cumulative distribution for the metric distances exponentiates into a familiar Sudakov form, but because of the two terms in \Eq{eq:tilde_s}, the probability of no emissions is controlled by the sum of the color Casimirs of the two jets, $C_A+C_B$.
Said another way, we must forbid emissions in \emph{both} jets to be larger than the observed value of $\tilde \ell_\beta$.
To double-logarithmic accuracy, we have
\begin{align}
\label{eq:double_log_distance}
\frac{d\varrho^{(\text{DL})}}{d\tilde \ell_\beta}= -\frac{2\alpha_s}{\beta\pi}(C_A+C_B)\frac{\log\tilde \ell_\beta}{\tilde \ell_\beta} e^{
-\frac{\alpha_s}{\beta\pi}(C_A+C_B)\,\log^2\tilde \ell_\beta}
\,.
\end{align}
The QCD color factors are $C_q=4/3$ for quark-initiated jets and $C_g = 3$ for gluon-initiated jets.  
This distribution between QCD jets of different origins was also calculated in \InRef{Komiske:2022vxg} for the original EMD, where it was interpreted as the dimension of the space of jets as a function of resolution or distance.
As shown in \Sec{sec:original_EMD_a1}, the original EMD and spectral EMD have the same behavior at double-logarithmic accuracy, so \Eq{eq:double_log_distance} holds in both cases.

This calculation illustrates the expected behavior of a metric distance.  
In general, gluon jets are more massive than quark jets because $C_g>C_q$, and as such the difference between the masses of gluon jets from one another is expected to be larger than for quark jets.  
Thus, the spectral EMD between two gluon jets is expected to be larger than the spectral EMD between two quark jets, as born out by this calculation.  
We will see in \Sec{sec:qgparton} that this same behavior is exhibited by parton shower simulations of quark- and gluon-initiated jets.

\section{Fixed-Order Correlations Between Jets at $e^+e^-$ Colliders}
\label{sec:fixedorder}

We now explore the structure of the spectral EMD through ${\cal O}(\alpha_s^2)$.
Our analysis will be based on hemisphere jets produced in $e^+e^-$ collisions, calculated at fixed order.
The leading non-trivial correlations between the two hemispheres are referred to as non-global logarithms \cite{Dasgupta:2001sh}, and we will be able to directly probe them through the spectral EMD distribution.
As a baseline, we compute the spectral EMD distribution between jets from distinct events, where such non-global effects are absent at this order.

\subsection{Isolating the Non-Trivial Correlations}

Up through ${\cal O}(\alpha_s^1)$, the spectral EMD between two jets is determined by their individual properties, with no non-trivial correlations.
Only at ${\cal O}(\alpha_s^2)$ do we see such correlations, so we would like to define an observable that isolates those effects.

We define the reduced spectral EMD as
\begin{equation}
\label{eq:reducedEMD}
\Delta_\beta(s_A, s_B) \equiv\text{SEMD}_\beta(\hat{s}_A, s_B) + \text{SEMD}_\beta(s_A, \hat{s}_B) - \text{SEMD}_\beta(\hat{s}_A, \hat{s}_B) - \text{SEMD}_\beta(s_A, s_B),
\end{equation}
where the reduced spectral functions are:
\begin{align}
\hat{s}_A(\omega) = E^2_A \, \delta(\omega), \qquad \hat{s}_B(\omega) = E^2_B \, \delta(\omega),
\end{align}
Because of the signs in \Eq{eq:reducedEMD}, larger values of $\Delta_\beta$ correspond to stronger correlations (i.e.~smaller distances) between jets. 

Furthermore, $\Delta_\beta \geq 0$ which we argue as follows.
Without loss of generality, we assume that $E_A \geq E_B$ and can then immediately write closed form expressions for much of $\Delta_\beta$.
First, the distance between reduced spectral functions is
\begin{align}
\text{SEMD}_\beta(\hat{s}_A, \hat{s}_B) = (E_A^2-E_B^2)\, \omega_{\max}, 
\end{align}
which is the cost of removing the excess squared energy from the origin to outside the jet.
Next, the distance between the reduced spectral function of jet $A$ and the full spectral function of jet $B$ is
\begin{align}
\text{SEMD}_\beta(\hat{s}_A, s_B) = \sum_{b,b'\in B} E_b \, E_{b'} \, \omega_{bb'} + (E_A^2-E_B^2) \, \omega_{\max},
\end{align}
because we must move all peaks in $s_B$ to the origin and remove the excess squared energy from the origin to outside the jet.
With these results so far, note that 
\begin{align}
\text{SEMD}_\beta(\hat{s}_A, s_B)-\text{SEMD}_\beta(\hat{s}_A, \hat{s}_B)  = \sum_{b,b'\in B}E_b E_{b'}\omega_{bb'} = \text{SEMD}_\beta(s_B,\hat{s}_B),
\end{align}
which is the distance between the reduced and full spectral functions of jet $B$ itself.
Then, $\Delta_\beta$ can be equivalently expressed as
\begin{align}
\Delta_\beta(s_A, s_B) =\text{SEMD}_\beta(s_B,\hat{s}_B) + \text{SEMD}_\beta(s_A, \hat{s}_B) - \text{SEMD}_\beta(s_A, s_B).
\end{align}
Because the spectral EMD is a metric, the triangle inequality holds, with
\begin{align}
\text{SEMD}_\beta(s_B,\hat{s}_B) + \text{SEMD}_\beta(s_A, \hat{s}_B) \geq \text{SEMD}_\beta(s_A, s_B) \geq 0,
\end{align}
and therefore $\Delta_\beta \geq 0$ as promised.

At ${\cal O}(\alpha_s^0)$, $\Delta_\beta$ is manifestly zero because $\hat{s}^{(0)}_A = s^{(0)}_A$ and $\hat{s}^{(0)}_B = s^{(0)}_B$.
Similarly, at  ${\cal O}(\alpha_s^1)$, where one of the jets must be massless, $\Delta_\beta$ is also zero.
Only at ${\cal O}(\alpha_s^2)$ do we get a non-zero reduced spectral EMD.
Following the same logic as in \Sec{sec:uptothree}, we find:
\begin{equation}
\label{eq:delta2}
\Delta^{(2)}_\beta(s_A, s_B) = \frac{1}{2}\sum_{\substack{a,a'\in J_A \\b,b'\in J_B}} \min\big[2E_a E_{a'}, 2 E_b E_{b'}, |E_B^2 - E_a^2 - E_{a'}^2|, |E_A^2 - E_b^2 - E^2_{b'}|\big] \, \min[\omega_{aa'},\omega_{bb'}],
\end{equation}
which holds even if $E_A \not= E_B$.
Note that this formula also works at lower orders, since $\Delta^{(2)}_\beta \to 0$ in the soft and/or collinear limits.

This expression for $\Delta_\beta$ has a similar structure to the minimum of the two jets' hemisphere energy correlation functions, with the key difference that the minimum is taken independently over the energy and angular factors.  
In general, the minimal energy and minimal angle of emission do not have to occur within the same jet.  
This has the interesting feature of significantly suppressing the contribution from one jet that has a soft, wide angle emission and the other jet with a hard, collinear emission.  
Effectively, to the observable $\Delta_\beta$, such a configuration would involve both a soft and a collinear emission.

For the subsequent calculations, we assume that the hemisphere jets come from events with a common center-of-mass energy $2E$, such that the Born-order hemisphere jet energy is $E$.
Because of energy-momentum conservation, the jet energies will in general differ from $E$ at higher orders. 
We define a dimensionless reduced spectral EMD as
\begin{equation}
\tilde{\Delta}_\beta \equiv \frac{\Delta_\beta}{E^2}.
\end{equation}
In \Sec{sec:fixed_order_double}, jets $A$ and $B$ correspond to hemisphere jets from \emph{different} events, and we randomly choose one jet per event.
In \Sec{sec:fixed_order_single}, jets $A$ and $B$ come from different hemispheres of the \emph{same} event.
By comparing these two distributions, we can isolate the effects of non-trivial correlations starting at ${\cal O}(\alpha_s^2)$.

\subsection{Distance Between Jets in Distinct Events to ${\cal O}(\alpha_s^2)$}
\label{sec:fixed_order_double}

Consider the reduced spectral EMD between hemisphere jets $A$ and $B$ in distinct uncorrelated events.
We randomly choose one hemisphere jet from each event, which ensures that $A$ and $B$ are identically distributed.
The distribution of the dimensionless reduced spectral EMD $\tilde{\Delta}_\beta$ is:
\begin{align}
\frac{d\varrho^{(\text{uncorr})}}{d \tilde \Delta_\beta} = \frac{1}{Z} \int d\vec x_A \int d\vec x_B\, |{\cal M}(\vec x_A)|^2\, |{\cal M}(\vec x_B)|^2\,\delta\left(\tilde \Delta_\beta - \frac{\Delta_\beta(s_A, s_B)}{E^2} \right),
\end{align}
where $Z$ is a normalization factor, $\vec x_A$ is the phase space coordinate for event $A$, $|{\cal M}(\vec x_A)|^2$ is the corresponding squared matrix element, and similarly for event $B$. 
Crucially, the spectral functions $s_A$ and $s_B$ are associated with the hemisphere jets (and not with the event as a whole).

By construction, this expression is only non-trivial starting at ${\cal O}(\alpha_s^2)$.
Expanding the squared matrix elements order by order, we obtain:
\begin{equation}
\label{eq:eejetuncorr}
\frac{d\varrho^{(\text{uncorr})}}{d\tilde \Delta_\beta} = \delta(\tilde \Delta_\beta) + \frac{1}{\sigma^2_0} \int d\vec x_A \int d\vec x_B\, |{\cal M}^{(1)}(\vec x_A)|^2\, |{\cal M}^{(1)}(\vec x_B)|^2\,\delta\left(\tilde \Delta_\beta-\frac{\Delta^{(2)}_\beta(s_A, s_B)}{E^2} \right) + \cdots,
\end{equation}
where $\sigma_0$ is the Born-order cross section, and $\Delta^{(2)}_\beta$ is defined in \Eq{eq:delta2}.
Terms proportional to $|{\cal M}^{(0)}(\vec x_A)|^2\, |{\cal M}^{(2)}(\vec x_B)|^2$ do not appear in this expression, since each jet requires at least two particles for $\tilde \Delta_\beta$ to be non-zero.
This distribution can be calculated numerically from the well-known matrix element for $e^+e^-\to q\bar q g$ scattering, as shown in \Sec{sec:numerical} below.

\subsection{Distance Between Jets in a Single Event to ${\cal O}(\alpha_s^2)$}
\label{sec:fixed_order_single}

Now consider the reduced spectral EMD between hemisphere jets $A$ and $B$ within the same event:
\begin{align}
\frac{d\varrho^{(\text{corr})}}{d\tilde \Delta_\beta} = \frac{1}{Z} \int d\vec x\, |{\cal M}(\vec x)|^2\,\delta\left(\tilde \Delta_\beta - \frac{\Delta_\beta(s_A, s_B)}{E^2} \right).
\end{align}
Though this expression is a proper cross section, and thus deserving of the symbol $\sigma$, we continue to use $\varrho$ for ease of comparison.

As with the uncorrelated case, this expression is only non-trivial starting at ${\cal O}(\alpha_s^2)$:
\begin{equation}
\frac{d\varrho^{(\text{corr})}}{d\tilde \Delta_\beta} = \delta(\tilde \Delta_\beta) + \frac{1}{\sigma_0} \int d\vec x\, |{\cal M}^{(2)}(\vec x)|^2\,\delta\left(\tilde \Delta_\beta-\frac{\Delta^{(2)}_\beta(s_A, s_B)}{E^2} \right)+\cdots,
\end{equation}
Here, we see the appearance of $|{\cal M}^{(2)}(\vec x)|^2$, since this is the first squared amplitude that allows each jet to have at least two particles each.
This distribution involves the matrix elements for $e^+e^-\to q\bar q gg$ and  $e^+e^-\to q\bar q q' \bar q'$, which we compute numerically using EVENT2 \cite{Catani:1996vz}.

\subsection{Numerical Results}
\label{sec:numerical}

We now show numerical results for the distributions of $\tilde{\Delta}_{\beta = 2}$ in the uncorrelated and correlated cases.  
We restrict to $\beta = 2$ for simplicity and familiarity with hemisphere jet masses.

\begin{figure}[t]
\begin{center}
\subfloat[]{
\includegraphics[width=0.45\textwidth]{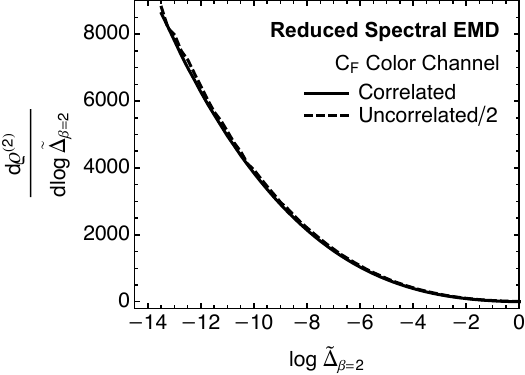}
\label{fig:as2plots_a}
}
$\qquad$
\subfloat[]{
 \includegraphics[width=0.45\textwidth]{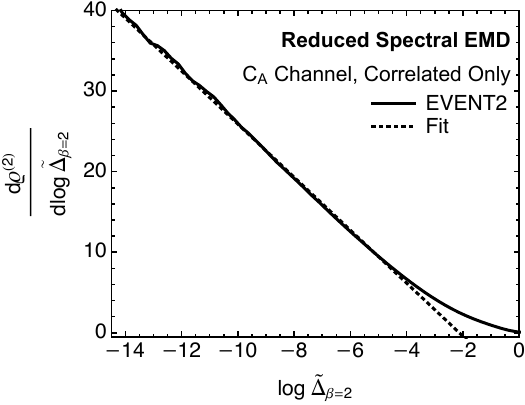}
 \label{fig:as2plots_b}
}\\
    \subfloat[]{
\includegraphics[width=0.45\textwidth]{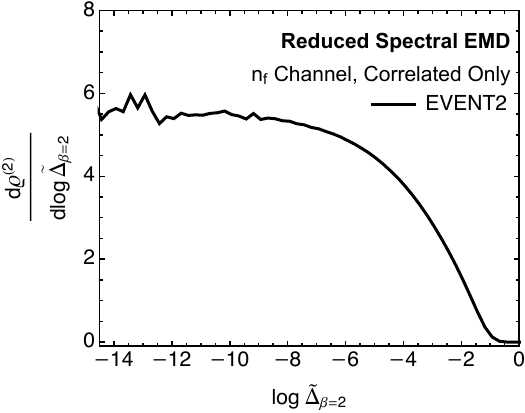}
\label{fig:as2plots_c}
}
\caption{
\label{fig:as2plots}
Distributions of the reduced spectral EMD distance $\tilde{\Delta}_{\beta = 2}$ calculated on jets production in $e^+e^-$ events at ${\cal O}(\alpha_s^2)$, separated by color channel.
For the $C_F$ channel in (a), the correlated distribution is a factor of 2 smaller than the uncorrelated one due to a Bose factor.
For the $C_A$ channel in (b), we compare the output of EVENT2 to a fit accounting for leading non-global logarithms.
For the $n_f T_R$ channel in (c), we simply show the output from EVENT2.
}
\end{center}
\end{figure}

In \Fig{fig:as2plots}, we show the distributions of $\log\tilde{\Delta}_{\beta = 2}$, where we remove overall factors of the coupling and color factors.  
That is, we plot $d\varrho^{(2)}/d\log \tilde \Delta_{\beta = 2}$, defined implicitly through
\begin{align}
\frac{1}{\sigma_0}\frac{d\varrho}{d\log \tilde \Delta_{\beta = 2}} = \frac{1}{\sigma_0}\left(
\frac{\alpha_s}{2\pi}
\right)^2C_F \, C^{(2)}\,\frac{d\varrho^{(2)}}{d\log \tilde \Delta_{\beta = 2}} + \cdots ,
\end{align}
where $\sigma_0$ is the Born-order cross section for $e^+e^-\to q\bar q$ scattering and $C^{(2)}$ is the appropriate color factor for the secondary emission.  
We study the three possible color structures:
\begin{equation}
C^{(2)} = \{C_F, C_A, n_f T_R\},
\end{equation}
where $n_f$ is the number of active quarks, and $T_R$ is the normalization of the fundamental representation of color SU(3).  

The $C^{(2)}=C_F$ channel appears for both the uncorrelated and correlated cases in \Fig{fig:as2plots_a}.  
Soft gluon emissions proportional to $C_F$ are emitted incoherently from one another, just like photons from charged particles.
Such emissions are described by the product of uncorrelated matrix elements $|{\cal M}^{(1)}|^2|{\cal M}^{(1)}|^2$.  
Two Abelian gluons produced in the same event are identical bosons, though, so there is a Bose factor of $1/2$ in the matrix element $|{\cal M}^{(2)}|^2$.  
Thus, in the deep infrared, where $\tilde \Delta_{\beta}\ll 1$, we expect that correlated contribution from $C_F$ gluon emission in $|{\cal M}^{(2)}|^2$ is a factor of 2 smaller than the uncorrelated contribution in $|{\cal M}^{(1)}|^2|{\cal M}^{(1)}|^2$.  
Indeed, the general trends of the distributions agree well, with differences arising at subleading order where specific angular or energy ordering becomes important.

The results for the $C^{(2)}=C_A$ channel are plotted in \Fig{fig:as2plots_b}, where only the correlated case contributes. 
The linear behavior in the deep infrared is expected from the form of leading non-global logarithms (NGLs).  
To obtain a non-zero value of $\tilde \Delta_{\beta}$, the two correlated gluons must be in different hemispheres, and therefore exhibit no collinear singularities, but can have hierarchical low energies.  
On this plot, we also include a linear fit and find that the leading logarithms (the slope on this plot) is well described by
\begin{align}
\frac{1}{\sigma_0}\frac{d\varrho^{(\text{corr})}}{d\log \tilde{\Delta}_{\beta = 2}}\supset -\frac{\pi^2}{3}\left(\frac{\alpha_s}{2\pi}\right)^2C_F \, C_A \log \tilde{\Delta}_{\beta = 2}\,,
\end{align}
as expected from the value of leading NGLs for hemisphere mass \cite{Dasgupta:2001sh}.  
Because there is no collinear singularity that contributes to the leading NGLs, the fact that the angular dependence of the hemisphere mass and $\Delta_{\beta = 2}$ are different has no effect.

Finally, we plot the $C^{(2)}=n_f \, T_R$ channel in \Fig{fig:as2plots_c}, where again only the correlated case contributes.   
There is no divergence associated with hierarchical energies from a gluon splitting into two quarks, so this distribution is only single logarithmic (flat on this plot) in the deep infrared.  
Along with the subleading logarithms in the $C_A$ channel, the fit we establish in these plots is
\begin{align}
\frac{1}{\sigma_0}\frac{d\varrho^{(\text{corr})}}{d\log\tilde\Delta_{\beta = 2}}\supset -7\left(\frac{\alpha_s}{2\pi}\right)^2C_F \, C_A+5.5\left(\frac{\alpha_s}{2\pi}\right)^2C_F \, n_f T_R\,.
\end{align}
By contrast, the values of the subleading hemisphere mass NGLs are \cite{Kelley:2011ng,Hornig:2011iu}:
\begin{align}
\frac{1}{\sigma_0}\frac{d\varrho^{(\text{sub-NGLs})}}{d\log\tilde\Delta_{\beta = 2}}&= \frac{3+18\zeta_3-11\pi^2}{9}\left(\frac{\alpha_s}{2\pi}\right)^2C_F \, C_A+\frac{4\pi^2-6}{9}\left(\frac{\alpha_s}{2\pi}\right)^2C_F \, n_f T_R\\
&\approx -9.3\left(\frac{\alpha_s}{2\pi}\right)^2 C_F \, C_A+3.7\left(\frac{\alpha_s}{2\pi}\right)^2C_F \, n_f T_R
\,.\nonumber
\end{align}
While close, we do not expect perfect agreement between subleading hemisphere mass NGLs and $\tilde \Delta_\beta$ because of the distinct energy and angular ordering in the definition of $\tilde \Delta_\beta$.

\section{Results from a Parton Shower}
\label{sec:qgparton}

Having established some resummed and fixed-order results, we now investigate spectral EMD distributions obtained from a parton shower.
We generate events in \textsc{MadGraph 3.4.0}~\cite{Alwall:2014hca} at a center-of-mass collision energy of 2 TeV, and use the following $\ell^+\ell^-\to jj$ processes to produce samples of quark and gluon jets:
\begin{itemize}
\item Quark jets: $e^+e^-\to u\bar u$;
\item Gluon jets: $\tau^+\tau^-\to gg$.\footnote{Collisions of tau leptons may seem unorthodox, but it is for simplicity within \textsc{MadGraph} so that the Higgs can mediate the production of gluon jets.}
\end{itemize}
These hard scattering events are showered using \textsc{Pythia 8.306} \cite{Bierlich:2022pfr} with its default settings, except when we turn off hadronization.  
Two exclusive $k_T$ jets \cite{Catani:1991hj} are found in each event with \textsc{FastJet 3.4.0} \cite{Cacciari:2011ma}, and one jet per event is chosen randomly for analysis.

\subsection{Distance between Jets at Parton Level}

\begin{figure}[t]
\begin{center}
\subfloat[]{
\includegraphics[width=0.45\textwidth]{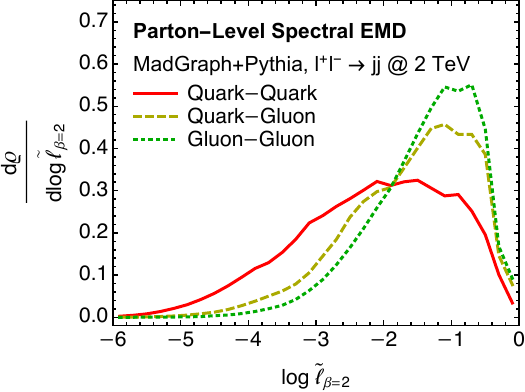}
\label{fig:nohadmet_a}
}
$\quad$
\subfloat[]{
\includegraphics[width=0.45\textwidth]{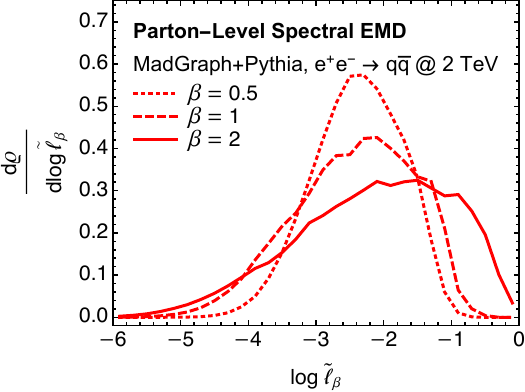}
\label{fig:nohadmet_b}
}
\caption{
\label{fig:nohadmet}
Parton-level distributions of the dimensionless spectral EMD $\tilde \ell_{\beta}$ for 2 TeV collisions.
(a) Comparison between two quark jets (solid, red), a quark and a gluon jet (dashed, yellow), and two gluon jets (dashed, green), for $\beta = 2$.
(b) Comparison between $\beta = \frac{1}{2}$ (dotted), $\beta = 1$ (dashed), and $\beta =2$ (solid), for the quark-quark sample. 
}
\end{center}
\end{figure}

Following \Eq{eq:tilde_s}, we compute the normalized spectral EMD $\tilde\ell_{\beta}$.
In this subsection, jets are simulated at parton level with no hadronization effects.

In \Fig{fig:nohadmet_a}, we plot the spectral EMD distribution between two quark jets, a quark and a gluon jet, and two gluon jets, focusing on $\beta = 2$.
There is a clear ordering to the distances between jets:
\begin{align}
\big\langle \log \tilde \ell_{\beta = 2} \big\rangle_{qq} <  \big\langle \log \tilde \ell_{\beta = 2} \big\rangle_{qg} <  \big\langle \log \tilde \ell_{\beta = 2} \big\rangle_{gg}\,,
\end{align}
where the subscripts denote the two jet categories being compared.  
From the double logarithmic analysis in \Eq{eq:double_log_distance}, the average log distance between jets is related to the sum of the color Casimirs:
\begin{align}
\label{eq:expval_dist}
\big\langle \log \tilde \ell_\beta \big\rangle_{AB} = - \frac{\pi}{2\sqrt{\alpha_s}} \frac{\sqrt{\beta}}{\sqrt{C_A+C_B}},
\end{align}
Taking ratios, this relation is well-reproduced by the \textsc{Pythia} parton-level samples:
\begin{align}
\frac{\big\langle \log \tilde \ell_{\beta=2} \big\rangle_{qq}}{\big \langle \log \tilde \ell_{\beta=2} \big\rangle_{qg}} &\approx 1.33\,, & \frac{\sqrt{C_F+C_A}}{\sqrt{2C_F}} &\approx 1.27\,,\\
\frac{\big\langle \log \tilde \ell_{\beta=2} \big\rangle_{qg}}{\big \langle \log \tilde \ell_{\beta=2} \big\rangle_{gg}} &\approx 1.16\,, & \frac{\sqrt{2C_A}}{\sqrt{C_F+C_A}} &\approx 1.18\,,\\
\frac{\big\langle \log \tilde \ell_{\beta=2} \big\rangle_{qq}}{\big \langle \log \tilde \ell_{\beta=2} \big\rangle_{gg}} &\approx 1.55\,, & \frac{\sqrt{C_A}}{\sqrt{C_F}} &= 1.5\,.
\end{align}

In \Fig{fig:nohadmet_b}, we compare the spectral EMD for angular weighting parameters $\beta = \frac{1}{2}$, $\beta = 1$, and $\beta = 2$, focusing on the quark-quark sample.
Here, we find:
\begin{align}
\big\langle \log \tilde \ell_{\beta = 1/2} \big\rangle_{qq} \approx  \big\langle \log \tilde \ell_{\beta = 1} \big\rangle_{qq}  \lesssim \big\langle \log \tilde \ell_{\beta = 2} \big\rangle_{qq}\,.
\end{align}
These mean ratios differ rather substantially from the leading logarithmic predictions (note the minus sign in \Eq{eq:expval_dist}):
\begin{align}
\frac{\big\langle \log \tilde \ell_{\beta=2}\big\rangle_{qq}}{\big\langle \log \tilde \ell_{\beta=1} \big\rangle_{qq}} &\approx 0.88\,, & \frac{\sqrt{2}}{\sqrt{1}} &\approx 1.41\,,\\
\frac{\big\langle \log \tilde \ell_{\beta=1}\big\rangle_{qq}}{\big\langle \log \tilde \ell_{\beta=\frac{1}{2}}\big\rangle_{qq}} &\approx 1.02\,, & \frac{\sqrt{1}}{\sqrt{0.5}} &\approx 1.41\,,\\
\frac{\big\langle \log \tilde \ell_{\beta=2}\big\rangle_{qq}}{\big\langle \log \tilde \ell_{\beta=\frac{1}{2}}\big\rangle_{qq}} &\approx 0.90\,, & \frac{\sqrt{2}}{\sqrt{0.5}} &= 2\,.
\end{align}
This difference seems to be due to physics at $\log \tilde\ell_\beta \approx 0$, where the double-logarithmic approximation is no longer accurate.
Importantly, at double-logarithmic accuracy, the maximal value of $\tilde\ell_\beta$ is 1, but the true upper bound depends on $\beta$ through the maximum angular value $\omega_{\max}  = 2^{\beta - 1}$.

We can reduce sensitivity to this upper bound effect by considering the variance of the spectral EMD.
Using \Eq{eq:double_log_distance}, the variance predicted from a double-logarithmic analysis is again related to the sum of the color Casimirs:
\begin{align}
\sigma_{\beta,AB}^2\equiv \big\langle \log^2 \tilde \ell_\beta \big\rangle_{AB}- \big\langle \log \tilde \ell_\beta \big\rangle_{AB}^2 = \frac{\pi(4-\pi)}{4 \alpha_s} \frac{\beta}{C_A+C_B}\,.
\end{align}
The ratios of the variances in the \textsc{Pythia} quark-quark sample are much better described than the means:
\begin{equation}
\frac{\sigma_{\beta=2,qq}^2}{\sigma_{\beta=1,qq}^2} \approx 1.71, \qquad \frac{\sigma_{\beta=1,qq}^2}{\sigma_{\beta=\frac{1}{2},qq}^2} \approx 1.83, \qquad \frac{\sigma_{\beta=2,qq}^2}{\sigma_{\beta=\frac{1}{2},qq}^2} \approx 3.13.
\end{equation}
Computing these ratios at higher orders would be an interesting avenue for future studies.

\subsection{Impact of Hadronization}

\begin{figure}[t]
\begin{center}
\subfloat[]{
\includegraphics[width=0.45\textwidth]{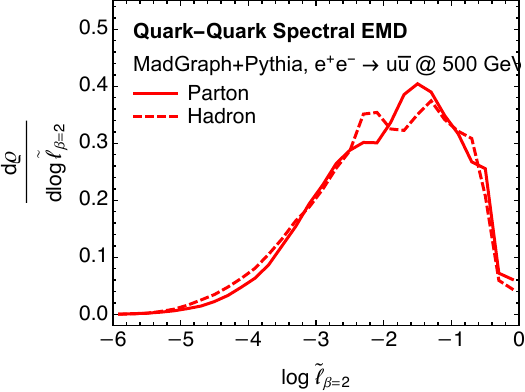}
\label{fig:hadmet_a}
}
$\quad$
\subfloat[]{
\includegraphics[width=0.45\textwidth]{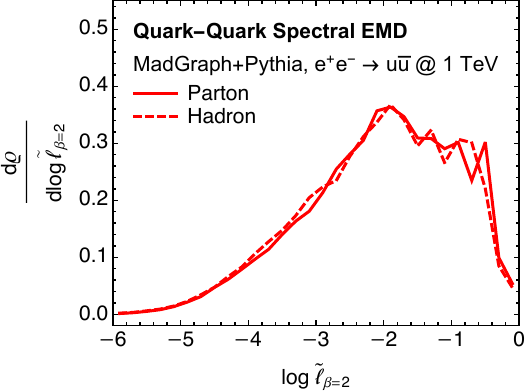}
\label{fig:hadmet_b}
}
\\
\subfloat[]{
\includegraphics[width=0.45\textwidth]{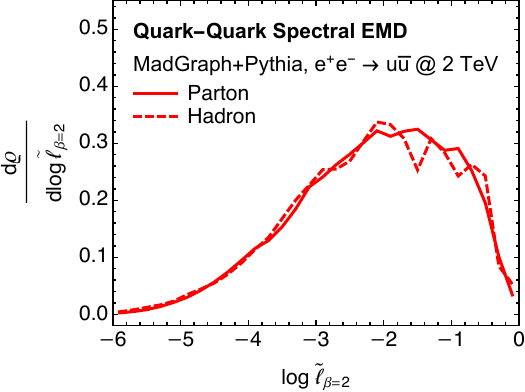}
\label{fig:hadmet_c}
}
\caption{
\label{fig:hadmet}
Comparison at parton-level (solid) versus hadron-level (dashed) for the dimensionless spectral EMD between two quark jets with $\beta = 2$.
Shown are three different center-of-mass collision energies: (a) 500 GeV, (b) 1 TeV, and (c) 2 TeV. 
}
\end{center}
\end{figure}

As a first study of the effect of non-perturbative physics on the spectral EMD, we compare the distributions of $\tilde \ell_{\beta = 2}$ between jets at parton-level versus hadron-level.
This is shown for quark jets ($e^+e^-\to u\bar u$) in \Fig{fig:nohadmet} for three different collision energies.
The difference between parton level and hadron level is relatively modest, with no discernible scaling with center-of-mass collision energy.

To try to understand if this non-perturbative insensitivity is expected, we can try to draw an analogy with jet mass.
As shown in \Sec{sec:two_particles}, the $\beta = 2$ spectral EMD at $\mathcal{O}(\alpha_s^1)$ is closely related to the sum of jet masses.
Perturbatively, the mass (or two-point energy correlation function) of a jet is well-understood \cite{Clavelli:1979md,Catani:1992ua,Banfi:2004yd}, which is closely related to the observable jet thrust and angularities \cite{Farhi:1977sg,Berger:2003iw,Ellis:2010rwa}.  
Because of its simplicity, the leading non-perturbative corrections to mass can be estimated by considering a jet with a single emission sensitive to the non-perturbative scale $\Lambda_\text{QCD} \simeq 1$ GeV, which is of comparable order to the Landau pole or hadron masses~\cite{Dokshitzer:1995zt}.
This non-perturbative emission has relative transverse momentum
\begin{align}
\Lambda_\text{QCD} \simeq E_{\rm NP} \, \theta_{\rm NP},
\end{align}
which depends on its energy $E_{\rm NP}$ and angle $ \theta_{\rm NP}$ from hard jet core.
The non-perturbative contribution to jet mass is dominated by wide-angle emissions with $\theta_{\rm NP} \simeq 1$, yielding
\begin{equation}
\delta_{\rm NP} \, m^2 \simeq E \, E_{\rm NP},
\end{equation}
where $E$ is the jet energy.

Consider $E = 250$ GeV, which is the approximate jet energy in \Fig{fig:nohadmet_a}.
If the analogy with jet mass held, then we would expect
\begin{align}
\delta_{\rm NP} \, \tilde \ell_{\beta = 2} \simeq \frac{\Lambda_\text{QCD}}{E}\simeq 0.004,
\end{align}
Note that $\log 0.004 \approx -5.5$, which is a region on this plot where the distribution has nearly vanished, so it is difficult to draw robust conclusions about what is happening given the small statistics.
Nevertheless, at $\log \tilde \ell_{\beta = 2} \sim -4$, the expected non-perturbative shift would push the distribution up to $\log \tilde \ell_{\beta = 2} \sim -3.8$.
If anything, the non-perturbative shift appears to go down by roughly this amount, in the opposite direction from the jet mass expectation.

One reason that the analogy with jet mass might be misleading is that the spectral EMD is not an additive observable.
An IRC-safe observable is additive if its value never decreases when a new emission is added to the jet \cite{Banfi:2004yd}, with jet mass being the canonical example.
For a large classes of additive observables, one can prove that non-perturbative physics affects the perturbative region via a simple positive shift of the distribution~\cite{Lee:2006fn}.
While the spectral EMD has an additive structure at ${\cal O}(\alpha_s)$, this no longer holds at higher orders.
Specifically, the value of the spectral EMD can decrease due to additional emissions at ${\cal O}(\alpha_s^2)$, because of the negative contribution appearing in \Eq{eq:genmetas2}.

Without an all-orders understanding of how the spectral EMD is modified by non-perturbative emissions, we cannot make more concrete statements at this point.
\InRef{Komiske:2019fks} showed how the EMD between parton- and hadron-level jets can be used to bound the non-perturbative shift on certain IRC-safe observables.
It is not clear, though, how to translate this into a bound on the shift of the spectral EMD distribution.
We leave a further analysis of non-perturbative effects to future work.

\section{Comparison with the Original EMD}
\label{sec:emdvsspec}

To gain more intuition for the spectral EMD, it is worth comparing its properties to the original EMD from \Eq{eq:original_EMD}.
Already at ${\cal O}(\alpha_s^0)$, the distances have quite different expressions due to the differing treatment of isometries.
For two jets $A$ and $B$ consisting of a single particle, the EMDs are:
\begin{align}
\text{SEMD}^{(0,0)}_{\beta}(s_A, s_B) &= |E_A^2-E_B^2|\,\omega_{\max},\\
\text{EMD}^{(0,0)}_{\beta}({\cal E}_A,{\cal E}_B) &= \max\big[E_A,E_B\big] \frac{\Omega_{AB}^\beta}{R^\beta} + \big| E_A - E_B \big|,
\end{align}
where $E_A$ and $E_B$ are the two jet energies and $\Omega_{AB}$ is their relative angle.
In addition to having different units, we see that the original EMD is sensitive to $\Omega_{AB}$ while the spectral EMD is not.

To do a meaningful comparison between the original EMD and spectral EMD, we assume that the two jets are aligned and have equal energies such that:
\begin{equation}
\Omega_{AB} = 0, \qquad E_A = E_B = E.
\end{equation}
With this assumption, both EMDs have values of $0$ at ${\cal O}(\alpha_s^0)$.
For the remainder of our analysis, we set $R = 1$ for simplicity.

For jets with more than one particle, we need to specify what we mean by $\Omega_{AB} = 0$.
We define the jet axis through a $\beta$-dependent weighted average of the jet constituents:
\begin{equation}
\label{eq:jet_axis}
\hat n_{A,\beta} = \argmin_{\hat{n}} \sum_{a \in J_A} \frac{E_a}{E_A} \big| \hat n - \hat{n}_a \big|^\beta,
\end{equation}
and similarly for $\hat n_B$.%
\footnote{This is an example of a recoil-free axis choice.  It would be interesting to study different axis choices here.  For example, some of the calculations below are much simpler using the winner-take-all axis~\cite{gavinwta,Bertolini:2013iqa,Larkoski:2014uqa}.}
For $\beta = 2$ and massless jet constituents, this corresponds to the usual jet axis aligned with the jet momentum:
\begin{equation}
\hat n_{A,\beta=2} =  \sum_{a \in J_A} \frac{E_a}{E_A} \hat{n}_a,
\end{equation}
Regardless of the value of $\beta$, we let
\begin{equation}
\label{eq:OmegaAB_align}
\Omega_{AB} = \Omega(\hat n_{A,\beta}, \hat n_{B,\beta}) \Rightarrow 0,
\end{equation}
which is implicitly $\beta$ dependent via the jet axes.

\subsection{${\cal O}(\alpha_s^1)$: Jets with Up to Two Particles}
\label{sec:original_EMD_a1}

Like in \Sec{sec:two_particles} at ${\cal O}(\alpha_s^1)$, one jet can have at most two particles, while the other jet must have one.
For the following analysis, let jet $A$ consist of particles $\{1,3\}$ and jet $B$ consist of particle $2$, with $E_1 + E_3 = E_2 = E$.

From \Eqs{eq:metas}{eq:lower_omega}, we see that the spectral EMD is
\begin{equation}
\text{SEMD}^{(1,0)}_{\beta}(s_A, s_B)  = E_1 E_3 \, \Omega_{13}^\beta.
\end{equation}
For the original EMD, all of the radiation from particles 1 and 3 has to be transported to particle 2, yielding:
\begin{align}\label{eq:emdexact}
\text{EMD}^{(1,0)}_{\beta}({\cal E}_A,{\cal E}_B)= E_1\,\Omega_{12}^\beta+E_3\, \Omega_{23}^\beta.
\end{align}
While the spectral EMD depends on the angle between particles in the same jet, the original EMD depends on the angle between particles in different jets.

For the special case of $\beta = 2$, we know from \Eq{eq:SEMD_mass} that the spectral EMD equals the squared jet mass at this order.
Taking the collinear limit $\Omega_{12} + \Omega_{23} \approx \Omega_{13}\ll 1$, the original EMD reduces the squared jet mass divided by the jet energy:
\begin{equation}
\text{EMD}^{(1,0)}_{\beta=2}({\cal E}_A,{\cal E}_B) \approx \frac{\text{SEMD}^{(1,0)}_{\beta}(s_A, s_B)}{E} = \frac{m_A^2}{E}\,,
\end{equation}
Note that thrust and $\beta = 2$ angularities \cite{Farhi:1977sg,Berger:2003iw,Ellis:2010rwa} are also proportional to the squared jet mass in the collinear limit. 
For generic values of $\beta$, we can solve \Eq{eq:jet_axis} in the collinear limit:
\begin{equation}
\Omega_{12} \approx \Omega_{13} \times
	\begin{cases}
	0 & 0< \beta \leq 1, \\
	\frac{\min[E_1, E_3]^{\frac{1}{\beta - 1}}}{E_1^{\frac{1}{\beta - 1}}+ {E_3^{\frac{1}{\beta - 1}}}} & \beta > 1,
	\end{cases}
\end{equation}
Taking also the soft limit of $E_3 \to 0$, we find: 
\begin{equation}
\text{EMD}^{(1,0)}_{\beta}({\cal E}_A,{\cal E}_B) \approx \frac{\text{SEMD}^{(1,0)}_{\beta}(s_A, s_B)}{E} \,. 
\end{equation}
Thus, up to normalization, the original and spectral EMDs agree at ${\cal O}(\alpha_s^1)$ in the simultaneous soft and collinear limits.
This means that the double-logarithmic analysis of \Sec{sec:dlogsec} holds for both cases, with differences appearing at higher orders.

Since jet $B$ consists of a single particle at this order, one way to interpret these EMDs is as the distance of closest approach between jet $A$ and the manifold of one-particle configurations.
For the original EMD, this interpretation was identified in \InRef{Komiske:2020qhg}, where it was shown more generally that the $N$-(sub)jettiness observables \cite{Brandt:1978zm,Stewart:2010tn,Thaler:2010tr,Kim:2010uj,Thaler:2011gf} corresponding to the distance of closest approach to $N$-particle manifolds.
This interpretation crucially relies on setting the jet axis via \Eq{eq:jet_axis}.
For the spectral EMD which is automatically invariant to isometries, the manifold of one-particle configurations consists of a single configuration with $s(\omega) = E^2 \delta(\omega)$.

\subsection{${\cal O}(\alpha_s^2)$: Jets with Up to Three Particles}

To the best of our knowledge, the expression for the original EMD at ${\cal O}(\alpha_s^2)$ has not been presented in the literature.
Just as in \Sec{sec:uptothree}, there are two phase space configurations to consider.
One contribution arises when jet $A$ has three particles $\{1,3,5\}$ and jet $B$ consists of a single particle $\{2\}$.
From \Eq{eq:spectral20}, the spectral EMD is:
\begin{equation}
\text{SEMD}^{(2,0)}_\beta(s_A, s_B)= E_1 E_3 \, \Omega_{13}^\beta+E_1E_5 \, \Omega_{15}^\beta +E_3E_5 \, \Omega_{35}^\beta.
\end{equation}
For the original EMD, we have:
\begin{align}
\text{EMD}^{(2,0)}_{\beta}({\cal E}_A,{\cal E}_B)= E_1\,\Omega_{12}^\beta+E_3\, \Omega_{23}^\beta + E_5\, \Omega_{25}^\beta.
\end{align}
Up to an overall energy scaling, these expressions agree in the strongly ordered limit with $E_1 \gg E_3 \gg E_5$ and $\Omega_{12} \ll \Omega_{23} \ll \Omega_{25}\ll 1$.

The second contribution arises when jet $A$ has two particles $\{1,3\}$ and jet $B$ also has two particles $\{2,4\}$.
From \Eq{eq:genmetas2}, the spectral EMD is:
\begin{equation}
\text{SEMD}^{(1,1)}_\beta(s_A, s_B) = E_1 E_3\Omega_{13}^\beta + E_2 E_{4} \Omega_{24}^\beta -  2\min[E_1E_3,E_2E_4] \min[\Omega_{13}^\beta,\Omega_{24}^\beta].
\end{equation}
The original EMD requires solving a genuine optimal transport problem in two dimensions.
Let $f \equiv f_{12}$ be the amount of energy transported from particle 1 to particle 2.
Because of the constraints in \Eq{eq:original_EMD_constraints}, the other elements of the transportation plan are fixed by $f$:
\begin{equation}
\label{eq:22minimization}
\text{EMD}^{(1,1)}_{\beta}({\cal E}_A,{\cal E}_B) = \min_{f} \Big[ f \, \Omega_{12}^\beta + (E_1 - f) \, \Omega_{14}^\beta + (E_2 -f) \, \Omega_{23}^\beta + (f - E_1 - E_2 + E) \,\Omega_{34}^\beta  \Big],
\end{equation}
where the energy coefficients must all be positive.

Solving the minimization in \Eq{eq:22minimization}, we find
\begin{align}
f &= \Theta\left(
\Omega_{12}^\beta-\Omega_{14}^\beta-\Omega_{23}^\beta+\Omega_{34}^\beta
\right)\max[0,E_1+E_2-E] \\
&
\hspace{3cm}+ \Theta\left(
-\Omega_{12}^\beta+\Omega_{14}^\beta+\Omega_{23}^\beta-\Omega_{34}^\beta
\right)\min[E_1,E_2] \,.\nonumber
\end{align}
The value of the original EMD depends on the hierarchy of the energies and angles.
We can express this as a minimization over the four possible hierarchies as: 
\begin{equation}\label{eq:EMD11sol}
\text{EMD}^{(1,1)}_{\beta}({\cal E}_A,{\cal E}_B) = \min_{\substack{a\neq a'\in J_A\\b\neq b'\in J_B\\E_a+E_b\leq E}}\left[
E_a\,\Omega_{ab'}^\beta + E_b\, \Omega_{a'b}^\beta + (E-E_a-E_b)\,\Omega_{a'b'}^\beta
\right]\,.
\end{equation}
Once again, the spectral EMD depends on the angle between particles in the same jet, while the original EMD depends on the angle between particles in different jets.

\begin{figure}
\begin{center}
\includegraphics[width=.4\textwidth]{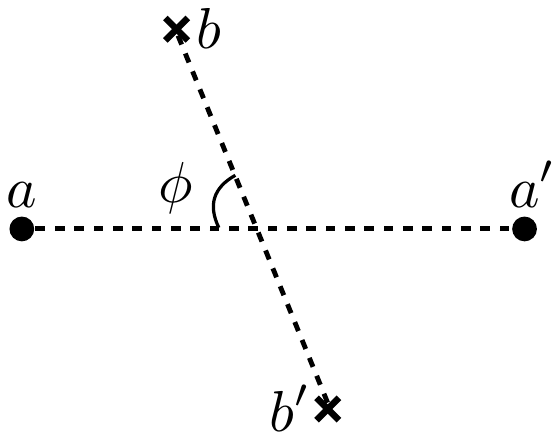}
\caption{\label{fig:emdangs}
Illustration of the azimuthal angle $\phi$ that quantifies the relative orientation of the two jets about their common axis in the collinear limit for calculating $\text{EMD}^{(1,1)}_{\beta=2}({\cal E}_A,{\cal E}_B)$.
}
\end{center}
\end{figure}

In the collinear limit for $\beta = 2$, we can express this EMD in a nice form that can we will further simplify in \Sec{sec:tangentEMD} by incorporating rotations about the jet axis.
As discussed above, to make a meaningful comparison between jets with the EMD, we align their axes, which for $\beta = 2$ means that we align their net momenta.
Then, we consider the particles in the two jets as illustrated in \Fig{fig:emdangs}, where the relative azimuthal angle $\phi$ of the particles in the two jets is measured between the common jet axis and particles $a$ and $b$.

Using the law of cosines in the collinear limit, the pairwise angles that appear in \Eq{eq:EMD11sol} can be expressed as:
\begin{align}
\Omega^2_{ab'}&\approx \frac{E_{a'}^2}{E^2}\Omega_{13}^2+\frac{E_b^2}{E^2}\Omega_{24}^2+2\frac{E_{a'}E_b}{E^2}\Omega_{13}\Omega_{24}\,\cos\phi\,,\\
\Omega^2_{a'b}&\approx \frac{E_{a}^2}{E^2}\Omega_{13}^2+\frac{E_{b'}^2}{E^2}\Omega_{24}^2+2\frac{E_{a}E_{b'}}{E^2}\Omega_{13}\Omega_{24}\,\cos\phi\,,\\
\Omega^2_{a'b'}&\approx \frac{E_{a}^2}{E^2}\Omega_{13}^2+\frac{E_b^2}{E^2}\Omega_{24}^2-2\frac{E_{a}E_b}{E^2}\Omega_{13}\Omega_{24}\,\cos\phi\,.
\end{align}
Then, in terms of the intrajet angles $\Omega_{13}$,$\Omega_{24}$ and the azimuth $\phi$, the EMD takes the simpler form:
\begin{align}\label{eq:emdcollb2}
\text{EMD}^{(1,1)}_{\beta=2}({\cal E}_A,{\cal E}_B) \approx  \min_{\substack{a\in J_A,\,b\in J_B\\E_a+E_b\leq E}}\frac{E_1E_{3}\Omega_{13}^2+E_2E_{4}\Omega_{24}^2+2E_a E_b\Omega_{13}\Omega_{24}\cos\phi}{E}\,.
\end{align}

\subsection{Incorporating Rotational Isometries}
\label{sec:tangentEMD}

For the above analysis, we assumed that we can freely translate the jets to align their axes, as in \Eq{eq:OmegaAB_align}.
As discussed in \InRef{deOliveira:2015xxd}, one can also perform rotations to further align the radiation.
This strategy of projecting the original EMD out by translations and rotations is known as the tangent EMD (TEMD) \cite{10.1007/978-3-642-40020-9_43}.
For general angles and $\beta$, we are unaware of a closed form expression for the TEMD.
By working in the collinear limit and fixing $\beta = 2$, though, we can gain insight from an approximate closed form expression.

Assuming that rotations about the jet axis are isometries, to calculate the TEMD in the collinear limit, we simply fix the relative azimuthal angle that appears in \Eq{eq:emdcollb2} to the value that minimizes the EMD.
This angle is clearly $\phi = \pi$, and so the TEMD in the collinear limit with $\beta = 2$ is
\begin{align}\label{eq:mintemd2}
\text{TEMD}^{(1,1)}_{\beta=2}({\cal E}_A,{\cal E}_B) \approx
\min_{\substack{a\in J_A,\,b\in J_B\\E_a+E_b\leq E}}\frac{E_1E_{3}\Omega_{13}^2+E_2E_{4}\Omega_{24}^2-2E_a E_b\Omega_{13}\Omega_{24}}{E}
\,.
\end{align}
From this expression, the original EMD differs by a non-negative term that depends on the relative azimuthal angle $\phi$:  
\begin{align}
\text{EMD}^{(1,1)}_{\beta=2}({\cal E}_A,{\cal E}_B) \approx  \text{TEMD}^{(1,1)}_{\beta=2}({\cal E}_A,{\cal E}_B) +
\max_{\substack{a\in J_A,\,b\in J_B\\E_a+E_b\leq E}}\frac{4E_a E_b\Omega_{13}\Omega_{24}}{E}\,\cos^2\frac{\phi}{2}\,.
\end{align}
Thus, we see that there is an explicit EMD cost to rotations about the jet axis, which enforces the relationship:
\begin{align}
\text{TEMD}^{(1,1)}_{\beta=2}({\cal E}_A,{\cal E}_B) \leq \text{EMD}^{(1,1)}_{\beta=2}({\cal E}_A,{\cal E}_B).
\end{align}

Both the TEMD and the SEMD respect isometries, so it is interesting to compare their behavior in the collinear limit with $\beta = 2$:
\begin{align}
\frac{\text{SEMD}^{(1,1)}_{\beta=2}(s_A, s_B)}{E} &= \text{TEMD}^{(1,1)}_{\beta=2}({\cal E}_A,{\cal E}_B)\\
&
\hspace{1cm}+\frac{2}{E}\biggl(\max_{\substack{a\in J_A,\,b\in J_B\\E_a+E_b\leq E}}E_a E_b\Omega_{13}\Omega_{24}-  \min[E_1E_3,E_2E_4] \min[\Omega_{13}^2,\Omega_{24}^2]\biggr)\,.\nonumber
\end{align}
The sign of the difference on the second line fixes the relative size of the SEMD with respect to the TEMD.
This difference can be either positive or negative, however, so there exists no fixed relationship between the SEMD and TEMD.
Considering the angular factors first, note that:
\begin{align}
\Omega_{13}\Omega_{24}\geq \min[\Omega_{13}^2,\Omega_{24}^2]\,.
\end{align}
For the energy factors, though, the hierarchy is the opposite:
\begin{align}
\max_{\substack{a\in J_A,\,b\in J_B\\E_a+E_b\leq E}}E_a E_b \leq \min[E_1E_3,E_2E_4]\,.
\end{align}
Therefore, the relative size of the SEMD and the TEMD depends in detail on how the energy is shared among the two particles in the jets and how that sharing compares to the relative angles between particles.
This also implies that the relative size of the original EMD and the SEMD depends sensitively on the particular momentum of the particles in the jets, which highlights the complementarity of these approaches.

\section{A Spectral Metric for Theory Space}\label{sec:thspacemet}

Finally, we introduce a spectral approach for constructing a metric on the space of theories.
As shown in \InRef{Komiske:2020qhg}, one can lift the original EMD to a cross-section mover's distance ($\Sigma$MD), which provides an data-driven way to define the distance between theories.
Here, we introduce the spectral $\Sigma$MD, which is invariant to the isometries of a theory by construction, even if the explicit form of those isometries is not known.

We note that the idea of a metric on theory space has a long history.
In conformal field theories (CFTs), the Zamolodchikov metric \cite{Zamolodchikov:1986gt,Kutasov:1988xb} is the canonical Riemannian metric on the space of theories, and it can be used to establish general and far-reaching results regarding the growth of degrees of freedom due to renormalization group evolution from high scales to low scales.
The differential Zamolodchikov metric, or line element, is defined as the value of a two-point correlation function between local operators constructed from tangent vectors on the space of theories.
This metric is therefore very nice for spaces that are smooth under variation of parameters, though not all CFTs are of this form.\footnote{See \InRef{Douglas:2010ic} for a discussion of issues and possible resolutions to the shortcomings of the Zamolodchikov metric.}
Furthermore, it is unclear how to practically use the Zamolodchikov metric to interpret realistic collider data.
More recent approaches to theory space include an application of information geometry in quantum field theory~\cite{Erdmenger:2020vmo} and a reformulation of  the exact renormalization group \cite{Polchinski:1983gv} as an optimal transport problem~\cite{Cotler:2022fze}.

In this section, we first review the original $\Sigma$MD before constructing its spectral variant.
For the 2-Wasserstein variant of the spectral $\Sigma$MD in particular, we can write down an explicit expression for the metric tensor in terms of Lagrangian parameters.
This metric tensor exhibits an intriguing link between the spectral $\Sigma$MD and renormalization group flow.

\subsection{Review of the Original $\Sigma$MD}

Consider a theory $\mathcal{T}$ defined as a set of events $\mathcal{E}_i$ with associated cross sections $\sigma_i$.
In analogy to \Eq{eq:energyflow}, we can define $\mathcal{T}$ as a distribution over events~\cite{Komiske:2020qhg}:
\begin{equation}
\mathcal{T}(\mathcal{E}) = \sum_{i=1}^N \sigma_i \, \delta(\mathcal{E} - \mathcal{E}_i).
\end{equation}
Integrating over all events, a theory is normalized via:
\begin{equation}
\int d \mathcal{E} \, \mathcal{T}(\mathcal{E}) = \sigma_{\rm tot},
\end{equation}
where $\sigma_{\rm tot}$ is the total cross section.
Strictly speaking, the volume element $d \mathcal{E}$ involves separate integrations over different multiplicity final states.

Given a distance between events $d(\mathcal{E}_i, \mathcal{E}_j)$, we can define a distance between theories as the work needed to rearrange theory $A$ to theory $B$.
Since the weight being moved is cross section, \InRef{Komiske:2020qhg} called this the cross-section mover's distance:
\begin{equation}
\text{$\Sigma$MD}_{\gamma,S;d}(\mathcal{T}_A,\mathcal{T_B}) = \min_{\mathcal{F}_{ab}} \sum_{a \in T_A} \sum_{b \in T_B} \mathcal{F}_{ab} \frac{d(\mathcal{E}_a,\mathcal{E}_b)^\gamma}{S^\gamma} + \bigg| \sum_{a \in T_A} \sigma_a - \sum_{b \in T_B} \sigma_b  \bigg|,
\end{equation}
where the transportation plan $\mathcal{F}_{ab}$ is constrained analogously to \Eq{eq:original_EMD_constraints}.
The $\Sigma$MD depends on $\gamma$ and $S$ (the analogies of $\beta$ and $R$ in the original EMD) and the choice of ground metric $d$ (the analogy of $\Omega$ in the original EMD).
While \InRef{Komiske:2020qhg} advocated setting $d(\mathcal{E}_i, \mathcal{E}_j)$ equal to $\text{EMD}_{\beta,R}(\mathcal{E}_i, \mathcal{E}_j)$, the $\Sigma$MD can be defined using any ground metric on event space.
The geometry on theory space is then induced by the $\Sigma$MD.

\subsection{Introducing the Spectral $\Sigma$MD}

Just as the spectral representation of an event in \Eq{eq:specfunc} is invariant to isometries, we can introduction a spectral representation of a theory that is invariant to isometries.
Unlike collision events or jets in particle physics, the isometries in theory space are not necessarily known {\it a priori}, unless one has a known model space being studied.
In the case of collision events, the solution was to represent an event exclusively in terms of pairwise angular distances, weighted by particle energies, which is clearly invariant to O(3) rotations or reflections of the celestial sphere.
In the case of theories, we can represent a theory in terms of pairwise distances between events, $\zeta(\mathcal{E}_i, \mathcal{E}_j)$.

With this motivation, we introduce the theory spectral function, which is defined through pairwise distances between events and weighted by event cross sections:
\begin{align}
{\mathfrak s}(\zeta) &= \sum_{i\in T} \sum_{j\in T} \sigma_i\sigma_j\, \delta\left(
\zeta-\zeta(\mathcal{E}_i, \mathcal{E}_j)
\right)\\
&=\sum_{i\in T}\sigma_i^2\delta(\zeta)+\sum_{i<j\in T} 2\sigma_i\sigma_j\, \delta\left(
\zeta-\zeta(\mathcal{E}_i, \mathcal{E}_j)
\right).\nonumber
\end{align}
Here, $T \equiv \{\mathcal{E}\}$ is a set of events as produced in some theory, $\zeta(\mathcal{E}_i, \mathcal{E}_j)$ is the distance between events $i$ and $j$, and $\sigma_i$ is the cross section for event $\mathcal{E}_i$.
By definition, pairwise distances are invariant to isometries, and any metric distance between events could be used to define $\zeta$, including the original EMD or spectral EMD.

From the theory spectral function, we can define its cumulative function as:
\begin{align}
{\mathfrak S}(\zeta) = \sum_{i\in T }\sigma_i^2\Theta(\zeta)+\sum_{i<j\in T} 2\sigma_i\sigma_j\, \Theta\left(
\zeta-\zeta(\mathcal{E}_i, \mathcal{E}_j)
\right)\,,
\end{align}
The spectral $\Sigma$MD between theory $A$ and theory $B$ is therefore:
\begin{align}\label{eq:thwasmet}
\text{S$\Sigma$MD}_{p=1}(\mathcal{T}_A, \mathcal{T}_B) \equiv \int d\zeta \, \left|{\mathfrak S}_A(\zeta)-{\mathfrak S}_B(\zeta)\right| ,
\end{align}
where the $p = 1$ subscript reminds us that this is a 1-Wasserstein distance.
We have suppressed integration bounds in this expression, but it ranges over all physical values of $\zeta$, from 0 up to the maximal distance between events $\zeta_{\rm max}$.

By construction, \Eq{eq:thwasmet} is a metric on the space of theory space spectral functions.
From the theorem of \InRef{BOUTIN2004709} (see \App{app:uniqueness}) we expect it to also be a metric on the space of theories modulo isometries, though there may potentially be some subtleties.
Technically, \InRef{BOUTIN2004709} proved that knowing the unordered pairwise distances of $n$ points in the space ${\mathbb R}^k$ is sufficient to uniquely determine the set of points, up to isometries, as long as all pairwise distances are distinct.
In general, we do not know what the manifold of theories is nor do we know its topology.
We expect, however, that this is not an issue by results like the Whitney \cite{whitney1936differentiable} (see, e.g., \Refs{prasolov2007elements,adachi2012embeddings} for modern presentations) or Nash \cite{nash1954c1,kuiper1955c1,kuiper1955c2} embedding theorems, which establish that smooth manifolds can be isometrically embedded in sufficiently high dimensional Euclidean space.
So, we assume that the theorem of \InRef{BOUTIN2004709} can be applied to generic theories as long as the number of events is countable, but we leave a more detailed justification or identification of limitations of this assumption to future work.

\subsection{Riemannian Theory Space}

A key advantage of the spectral $\Sigma$MD over the original $\Sigma$MD is that we can express it in closed form.
The 2-Wasserstein metric has a Riemannian structure \cite{lott2006ricci,lott2007geometric,villani2008optimal} and this can be used to extract a metric tensor for theory space, expressed in terms of the cumulative theory spectral functions.

Analogous to \Eq{eq:SEMD_p_def}, the (squared) 2-Wasserstein version of \Eq{eq:thwasmet} is:
\begin{align}\label{eq:riemannthmet}
\text{S$\Sigma$MD}_{p=2}(\mathcal{T}_A, \mathcal{T}_B) \equiv \int d\sigma^2 \, \left|{\mathfrak S}_1^{-1}(\sigma^2)-{\mathfrak S}_2^{-1}(\sigma^2)\right|^2, 
\end{align}
where ${\mathfrak S}^{-1}(\sigma^2)$ is the inverse cumulative spectral function, whose argument is a squared cross section.
With continuous distributions of pairwise event distances, we can express this (squared) Riemannian metric in terms of the cumulative spectral function.
First, the cumulative spectral function is
\begin{align}
\mathfrak{S}(\zeta) = \int d\sigma_i\,d\sigma_j\, \Theta\left(\zeta - \zeta(\mathcal{E}_i, \mathcal{E}_j) \right),
\end{align}
where $d\sigma_i$ is the differential cross section squared for events in ensemble $i$.
The inverse cumulative spectral function can be written as
\begin{align}
\label{eq:invthspec}
\mathfrak{S}^{-1}(\sigma^2) = \int d\zeta\,\Theta\left(
\sigma^2 - \mathfrak{S}(\zeta)
\right)\,.
\end{align}
This expression satisfies the inverse function theorem with
\begin{align}
\frac{d\mathfrak{S}^{-1}}{d\sigma^2} = \left(\frac{d\mathfrak{S}}{d\zeta} \right)^{-1}.
\end{align}
The expression in \Eq{eq:invthspec} can be used to evaluate the $p = 2$ spectral $\Sigma$MD of \Eq{eq:riemannthmet}.

To convert the spectral $\Sigma$MD into a metric tensor, we need to define coordinates on theory space.
Let the theory spectral functions be dependent on a set of parameters $\{\lambda\}$, like couplings or masses, that define the theory.
We will be interested in the spectral $\Sigma$MD between a theory evaluated at energy scales $Q$ and $Q+dQ$, respectively, where these parameters change under renormalization group flow as
\begin{align}
\beta_i = Q\frac{\partial\lambda_i}{\partial Q}.
\end{align}
Assuming that $Q$ only appears implicitly through the parameters $\{\lambda\}$, the inverse cumulative spectral function transforms as:
\begin{equation}
{\mathfrak S}^{-1}(\sigma^2;\{\lambda(Q+dQ)\}) = {\mathfrak S}^{-1}(\sigma^2;\{\lambda (Q)\}) + \sum_i \beta_i  \frac{d {\mathfrak S}^{-1}(\sigma^2;\{\lambda (Q)\})}{d \lambda_i}  \frac{dQ}{Q}+\cdots\,.
\end{equation}
Plugging this into \Eq{eq:riemannthmet}, the differential line element is:
\begin{align}
\label{eq:line_element}
\lim_{dQ\to 0} \sqrt{\text{S$\Sigma$MD}_{p=2}\big(
{\mathfrak s}(\zeta;\{\lambda(Q+dQ)\}),{\mathfrak s}(\zeta;\{\lambda(Q)\})
\big)} =\bigg(\sum_{i,j} \beta_i\beta_j g_{ij} \bigg)^{1/2} \frac{dQ}{Q},
\end{align}
where the symmetric rank-two tensor $g_{ij}$ can be viewed as a metric on theory space:
\begin{align}
\label{eq:gij_one_way}
g_{ij} \equiv \int d\sigma^2 \, \frac{\partial{\mathfrak S}^{-1}(\sigma^2;\{\lambda\})}{\partial\lambda_i}\,\frac{\partial{\mathfrak S}^{-1}(\sigma^2;\{\lambda\})}{\partial\lambda_j} .
\end{align}

Since it is often more convenient to compute the cumulative spectral function $\mathfrak{S}(\zeta)$ rather than its inverse $\mathfrak{S}_1^{-1}(\sigma^2)$, we provide an alternative expression for the metric tensor $g_{ij}$.
Taking derivates of \Eq{eq:invthspec} with respect to $\lambda_i$, we find:
\begin{align}
\frac{\partial{\mathfrak S}^{-1}}{\partial\lambda_i} = -\left(\frac{d\mathfrak{S}}{d\zeta} \right)^{-1} \frac{\partial\mathfrak S}{\partial\lambda_i}.
\end{align}
Using the fact that the cross-section-squared coordinate is related to the event distance coordinate $\zeta$ as $\sigma^2 = \mathfrak{S}(\zeta)$, we find:
\begin{align}
d\sigma^2 = \frac{d\mathfrak{S}}{d\zeta}\,d\zeta\,.
\end{align}
Inserting these relations into \Eq{eq:gij_one_way} yields an alternative form for $g_{ij}$:
\begin{align}
\label{eq:gij_two_way}
g_{ij} = \int d\zeta\, \left({\frac{d\mathfrak{S}(\zeta;\{\lambda\})}{d\zeta}}\right)^{-1} \frac{\partial \mathfrak{S}(\zeta;\{\lambda\})}{\partial \lambda_i}\,\frac{\partial \mathfrak{S}(\zeta;\{\lambda\})}{\partial \lambda_j}.
\end{align}

The differential line element in \Eq{eq:line_element} shares a key property with the Zamolodchikov metric, namely that non-zero distances are only accumulated with non-zero $\beta$-functions.
A key difference, though, is that the spectral $\Sigma$MD has a direct connection to measured quantities in collider events.
Here, we just note this fascinating connection and leave a deeper interpretation and understanding to future work.

\section{Conclusions}
\label{sec:concs}

Equipping the space of particle collision data with a metric opens up a suite of geometric data analysis strategies.
The spectral EMD introduced in this paper offers a complementary approach to the original EMD from \InRef{Komiske:2019fks}.
The spectral EMD respects isometries by construction, unlike the original EMD which is faithful (though not invariant) to the symmetries of the ground metric.
Futhermore, since spectral functions are one-dimensional objects, we can avoid the numerical optimization needed for two-dimensional optimal transport.
Two drawbacks of the spectral approach is that not every spectral function corresponds to a physical arrangement of particles, and when it does, the spectral function redundantly encodes the particle information.
We view these as reasonable tradeoffs to achieve closed form expressions for the spectral EMD that are amenable to  precision calculations.

This paper has just scratched the surface of potential applications and consequences of the spectral EMD, and there are several questions introduced in the text that deserve further study.
Both the original EMD and spectral EMD define a metric space for collider events; can their similarities and differences be made more precise and quantitative?
Both the tangent EMD and spectral EMD are invariant to isometries and have the same behavior at double logarithmic accuracy; how do their properties differ going to higher orders?
The spectral EMD is not an additive observable, which complicates the analysis of non-perturbative effects; can we nevertheless understand the apparently small impact of hadronization?
The spectral function provides a unique jet representation up to isometries and sets of measure zero; is there an experimental impact from degeneracies that can appear due to finite angular resolution?
We focused on a spectral EMD construction based on the 1-Wasserstein metric where there is a duality between energy ordering and angular ordering; are there advantages from instead using the 2-Wasserstein metric which exhibits a Riemannian structure?
Going to theory space, we found an intriguing connection between the spectral $\Sigma$MD and the Zamolodchikov metric; can this relation be sharpened, and what distinguishes theories at different points along their renormalization group flow?

We hope that the spectral EMD and the way that the spectral function encodes information finds broad applications for physics analyses.
For example, the spectral function approach may provide a novel method for extracting physical quantities like the strong coupling $\alpha_s$ or the top quark mass $m_t$.
With the connection to a theory space metric, perhaps it would unlock new ways to observe and measure the QCD $\beta$-function, through the flow of QCD as different energy scales are probed.
This could also provide a new perspective on entropy growth as a parton shower evolves to the infrared \cite{Neill:2018uqw,Cheung:2023hkq}.
Recently, a procedure for measuring the top quark mass was proposed that exploited the correlation between pairwise angles and mass scales in the hadronic decay of a top quark \cite{Holguin:2022epo}, utilizing three-point energy correlators.
Energy correlators are the first moment of the spectral function, or its higher-point generalizations, and their structure has an immediate interpretation as the correlation function of local operators on the celestial sphere.

Finally, the presence of a (Riemannian) metric implies that collider data lives on a manifold, and the properties of this manifold can be studied.
The study of \InRef{Larkoski:2020thc} showed that $N$-body massless relativistic phase space is the product of a simplex and a hypersphere, which has a non-trivial topology as encoded in homotopy groups.  
This non-trivial topology has consequences for machine learning on the data and may present obstructions that cannot be overcome by some architectures \cite{Batson:2021agz}.
The collider geometry induced by a metric is sensitive to the structure of phase space and four-momentum conservation, but also implicitly involves the squared matrix element.
In perturbation theory, matrix elements generically exhibit divergences at phase space boundaries, which might dramatically alter the geometry and topology relative to the naive expectation from phase space.
Does the simple form of the spectral EMD enable the prediction, calculation, and observation of quantities like the Ricci curvature in data as an optimal transport space \cite{lott2009ricci}?
We hope that answers to these questions and more will produce a rich and fruitful perspective on the vast quantities and high dimensionality of collider physics data.

\begin{acknowledgments}

We thank Rikab Gambhir for comments on the manuscript and Nathaniel Craig for suggesting comparing to the 2-Wasserstein distance.
This work was supported in part by the UC Southern California Hub, with funding from the UC National Laboratories division of the University of California Office of the President.
J.T. was supported by the U.S. Department of Energy (DOE) Office of High Energy Physics under Grant Contract No. DE-SC0012567, by the DOE QuantISED program through the theory consortium ``Intersections of QIS and Theoretical Particle Physics” at Fermilab (FNAL 20-17), and by the National Science Foundation under Cooperative Agreement No. PHY-2019786 (The NSF AI Institute for Artificial Intelligence and Fundamental Interactions, \url{http://iaifi.org/}).

\end{acknowledgments}

\appendix

\section{Uniqueness of the Spectral Representation}
\label{app:uniqueness}

As mentioned in \Sec{sec:spectralfunctionproperties}, the spectral function determines an event uniquely up to isometries and configurations of measure zero.
In this appendix, we present a proof of this statement and highlight some pathological cases.

It is worth mentioning that there are other approaches to enforce isometries.
For example, one can construct irreducible representations of the isometry group from invariant polynomials of particle momenta.
This is similar in spirit to efforts to establish invariant operator bases on which equations of motion redundancies have been eliminated; see, e.g., \InRef{Henning:2015daa}.
The systematic procedure for doing this is rather challenging, though, especially as the multiplicity of operators (or particles) increases. 
Alternatively, if one is willing to forgo irreducibility, then one can use a jet representation that is overcomplete, like $C$-correlators \cite{Tkachov:1995kk} or energy flow polynomials \cite{Komiske:2017aww}.
Such bases can be excessively overcomplete, though, requiring tens of thousands of terms, even for modest particle multiplicities.%
\footnote{See \Refs{Komiske:2019asc,Cal:2022fnm} for some techniques to mitigate redundancies.}

By contrast, the spectral function of a jet with $n$ resolved particles has ${n\choose 2}+1$ $\delta$-functions, corresponding to all distinct particle pairs and the contact term at $z=0$.
While some of this information is redundant, it is not excessively so, with a quadratic growth of the required information with multiplicity $n$.
Crucially, this information is encoded in such a way that distinct jets are represented distinctly, up to isometries.

\subsection{Proof of Uniqueness}

The key assumption for the following proof is that all pairwise distances between particles in a jet are distinct real numbers.
For experimental data, particle locations in the tracker or calorimeter are discrete because of the finite angular resolution of the detectors.
This complicates the construction of the jet from its spectral function as presented here, but we leave a detailed study to future work.

The proof of the uniqueness of the spectral function representation goes as follows.
We start with angular information.
The spectral function clearly encodes all pairwise distances between particles in the jet through the location of its $\delta$-function spikes.
These pairwise distances are real numbers and in general have continuous, non-zero probability to lie anywhere in the interval $\omega \in [0,\omega_{\rm max}]$.
A jet only ever has a finite integer multiplicity, and so there is strictly 0 probability that two distinct pairs of particles have the same pairwise distance.

With this, we can apply Theorem 2.6 of \InRef{BOUTIN2004709}, which established the conditions for constructibility of $n$ points on the plane from their distribution of pairwise distances.
They prove the following result.
Let $P=(p_1,p_2,\dotsc,p_n)$ and $Q=(q_1,q_2,\dotsc,q_n)$ be two collections of points on the plane, each drawn from some non-degenerate, continuous probability distributions.
Then, with probability 1, if the set of unordered pairwise distances of points in $P$ and $Q$ are identical, then there exists an isometry in $\text{E}(2)\otimes S_n$ that renders the sets $P$ and $Q$ identical.\footnote{The theorem of \InRef{BOUTIN2004709} is actually much more general and establishes equality of collections $P$ and $Q$ of points in $\mathbb{R}^k$ up to the action of an arbitrary isometry group, if their distributions of unordered pairwise distances are identical.}
Because we assume 0 probability for multiple pairwise distances between particles in a jet to be identical, this theorem establishes that the location of particles in a jet on the celestial sphere can be reconstructed uniquely, up to isometries, with probability 1.

We next turn to energy information.
To establish particle $i$'s energy, consider the three distinct particles $i,j,k$ in the jet.
By the result above, the distances between particle pairs $ij$, $ik$, and $jk$ are distinct, and distinct from any other pairwise distances.
Therefore, we can identify them in the spectral function.
The weights $w$ with which these pairwise distances occur in the spectral function is the product of the corresponding energies:
\begin{align}
w_{ij} = 2E_iE_j, \qquad w_{ik} = 2E_iE_k, \qquad w_{jk} = 2E_jE_k.
\end{align}
It then follows that the energy of particle $i$ is:
\begin{align}
E_i = \sqrt{\frac{w_{ij}w_{ik}}{2w_{jk}}}\,.
\end{align}
This procedure can be continued for all particles in the jet.
Therefore, both the spatial distribution of the particles in the jet and their energies can be (nearly) uniquely reconstructed from the spectral function, up to isometries.

\subsection{Near-Degenerate Configurations}

We argued above that the spectral function uniquely specifies a jet with probability 1.
There is, however, a non-empty set of spectral functions that do not uniquely encode the location of particles, though this set has non-zero codimension with respect to the full space of jets.
Nevertheless, there may be nearly degenerate subspaces with non-zero probability that could introduce significant ambiguities.
One way these subspaces could arise is if the particles are confined to a grid with finite resolution, as in any realistic experimental detector.  
Here, we look at an example of a near-degenerate configuration that might require a more careful treatment.

Consider two jets, $A$ and $B$, with the same total energy.
Let jet $A$ consist of two particles $\{1,3\}$, with spectral function:
\begin{equation}
s_A(\omega)=(E_1^2+E_3^2 )\delta(\omega)
+ 2E_1E_3\,\delta( \omega-\omega_{13} ).
\end{equation}
Let $B$ consist of three particles $\{2,4,6\}$, with spectral function:
\begin{equation}
s_B(\omega) =(E_2^2+E_4^2+E_6^2 )\delta(\omega)
+ 2E_2 E_4\,\delta(\omega - \omega_{24})
+ 2E_2 E_6\,\delta(\omega - \omega_{26} )
+ 2E_4 E_6\,\delta(\omega - \omega_{46} ).
\end{equation}
For general $\theta_{24}$, $\theta_{26}$, and $\theta_{46}$, $s_B(\omega)$ has three peaks for $\omega>0$, so it can clearly be distinguished from $s_A(\omega)$, which only has one.
As the particles in jet $B$ approach an equilateral triangle configuration, though, the three peaks of $s_B(\omega)$ degenerate to a single peak.
If $\theta_{13} = \theta_{24}=\theta_{26}=\theta_{46}$, then $s_B(\omega)$ will have a single peak at the same location as for $s_A(\omega)$.
Furthermore, these peaks will have the same height if $E_1 E_3 =E_2 E_4+ E_2 E_6 + E_4 E_6$.
Thus, to render these spectral functions identical, there must be four constraints imposed on the structure of $s_B(\omega)$, such that the degenerate subspace has codimension 4.

To estimate when experimental resolution might render this configuration problematic, assume that there is some angular resolution $\epsilon_\theta$ within which all angles are equal and an energy resolution $\epsilon_E$ within which energies are equal.
Then, the probability that these two spectral functions are within resolution of one another scales like $\epsilon_\theta^3 \epsilon_E$, assuming continuous and smooth probability distributions for pairwise angles and particle energies.
A conservative estimate of these resolution factors exclusively from the resolution of calorimetry at ATLAS or CMS at the Large Hadron Collider, for example, assumes $\epsilon_\theta\sim \epsilon_E\sim 0.1$, so these nearly degenerate configurations are suppressed by at least $10^{-4}$ with respect to generic particle momenta, with no other assumptions on the distributions of particles.
There may be applications that enhance these nearly-degenerate contributions, for example if there are low multiplicity configurations from resonance decays that have a preferred angular scales.
The main lesson from this study is that one has to be mindful of configurations that can be close in SEMD even if they are far apart in EMD.

\section{Comparison with the 2-Wasserstein Metric}
\label{sec:riemannspec}

In the body of this paper, we focused on the 1-Wasserstein metric for defining the spectral EMD.
As discussed in \Eq{eq:SEMD_p_def}, it is also natural to define the ($p$-th power of) the $p$-Wasserstein metric.
The $p = 2$ case is particularly interesting, since the 2-Wasserstein distance enjoys many of the properties of a Riemannian metric, like the uniqueness of (affine) geodesics.
Indeed, Riemannian manifolds like the surface of the Earth, or Lorentzian manifolds like space-time, are the most familiar contexts in which a metric is used in physics.
In this appendix, we briefly discuss some properties of the 2-Wasserstein metric on the spectral function, leaving an in-depth study to future work.

Following \Eq{eq:SEMD_p_def} the (squared) 2-Wasserstein distance between two spectral functions $s_A(\omega)$ and $s_B(\omega)$ is:
\begin{equation}
\label{eq:SEMD_2_def}
\text{SEMD}_{\beta,p=2}(s_A, s_B) \equiv \int_{0}^{E^2} d\tilde{E}^2\, \big| S^{-1}_A(\tilde{E}^2) - S^{-1}_B(\tilde{E}^2) \big|^2 .
\end{equation}
For simplicity, we have assumed that the two jets have the same total energy $E$.
Unlike the $p = 1$ case from \Eq{eq:SEMD_def}, where we could define the spectral EMD through the cumulative spectral function, the $p = 2$ case only has a closed form expression in terms of the inverse cumulative spectral function $S^{-1}$, whose argument is a squared energy and whose value is a pairwise angular distance.

Now we want to repeat parts of the analysis of \Sec{sec:multcalc} and consider the $p = 2$ spectral EMD on low-multiplicity jets.
If two jets $A$ and $B$ each consist of a single particle with the same energy $E$, then their $p = 2$ spectral EMD is zero:
\begin{equation}
\text{SEMD}^{(0,0)}_{\beta,p=2}(s_A, s_B) = 0.
\end{equation}
The first non-trivial configuration is when jet $A$ has two particles $\{1,3\}$, with spectral function given in \Eq{eq:specA_1}, and jet $B$ still has one particle.
The corresponding inverse cumulative distributions, on the domain $\tilde{E}^2\in[0,E^2]$, are:
\begin{align}
S_A^{-1}(\tilde{E}^2)&=\Theta(\tilde{E}^2-E_1^2-E_3^2) \, \omega_{13},\\
S_B^{-1}(\tilde{E}^2)&= 0.
\end{align}
This yields a $p = 2$ spectral EMD of
\begin{align}
\text{SEMD}^{(1,0)}_{\beta,p=2}(s_A, s_B) &= \int_0^{E^2} d\tilde{E}^2\, \omega_{13}^2\,\Theta(\tilde{E}^2-E_1^2-E_3^2) = 2E_1E_3 \omega_{13}^2.
\end{align}
Using \Eq{eq:lower_omega} with $\beta = 1$ (not $\beta = 2$), this correspond to half of the squared jet mass.
As a rule of thumb, spectral EMDs with the same value of $\beta p$ have similar low-order behaviors.

We can continue and calculate the distance between two jets each with two constituents.
This is the situation studied in \Sec{sec:example_ot}, where the inverse cumulative spectral functions are given in \Eqs{eq:inv_cum_A}{eq:inv_cum_B}.
This yields a $p = 2$ spectral EMD of
\begin{align}
\text{SEMD}^{(1,1)}_{\beta,p=2}&=
\int_0^{E^2} d\tilde{E}^2 \left|\omega_{13}\,\Theta(\tilde{E}^2-E_1^2-E_3^2)-\omega_{24}\,\Theta(\tilde{E}^2-E_2^2-E_4^2)\right|^2\\
&=
2E_1E_3\omega_{13}^2+2E_2E_4\omega_{24}^2-\,4\min[E_1E_3,E_2E_4]\omega_{13}\omega_{24}.\nonumber
\end{align}
The $p=2$ spectral EMD is similar to the $p = 1$ case from \Eq{eq:genmetas2} in form, but note that the last subtracted factor involves a product of pairwise angles in the two jets, rather than their minimum.

\bibliography{metric}

\end{document}